\documentclass[12pt,preprint]{aastex}
\newcommand       \be           {\begin{equation}}
\newcommand       \ee           {\end{equation}}

\newcommand       \K            {\,{\rm K}}
\newcommand       \ab            {{\,{\rm a}}}
\newcommand       \g            {\,{\rm g}}
\newcommand       \m            {\,{\rm m}}
\newcommand       \mum           {\,\mu{\rm m}}
\newcommand       \mic           {\,\mu{\rm m}}
\newcommand       \cm           {\,{\rm cm}}
\newcommand       \km           {\,{\rm km}}
\newcommand       \s            {\,{\rm s}}
\newcommand       \erg          {\,{\rm erg}}
\newcommand       \kms		{\,{\rm km \, s}^{-1}}
\newcommand	  \pc		{\,{\rm pc}}
\newcommand	  \yr		{\,{\rm yr}}
\newcommand	  \Myr		{\,{\rm Myr}}
\newcommand       \nH           {n_{\rm H}}

\newcommand       \Qabs	        {Q_{\rm abs}}

\newcommand       \gtsim        {\gtrsim}
\newcommand       \ltsim        {\lesssim}
\newcommand	  \E	        {\erg \cm^{-3}}
\newcommand	  \vej		{v_{\rm ej}}

\shorttitle{Extra-Solar Dust}
\shortauthors{Murray Weingartner \& Capobianco}

\begin{document}

\title{On the Flux of Extra-Solar Dust in Earth's Atmosphere} 

\author{N. Murray\altaffilmark{1}, Joseph C. Weingartner \& P. Capobianco\altaffilmark{2}}
\affil{CITA, 60 St. George Street, University of
Toronto, Toronto, ON M5S 3H8, Canada}
\email{murray@cita.utoronto.ca, weingart@cita.utoronto.ca}
\altaffiltext{1}{Canada Research Chair in Astrophysics}
\altaffiltext{2}{Current address: Department of Physics and Astronomy,
  University of Victoria, Elliott Building, 3800 Finnerty Rd.,
  Victoria, BC, V8P 1A1, Canada}

\begin{abstract}
Micron size extrasolar dust particles have been convincingly detected
by satellites. Larger extrasolar meteoroids (5-35 microns) have most
likely been detected by ground based radar at Arecibo and New
Zealand. We present estimates of the minimum detectable particle sizes
and collecting areas for both radar systems. We show that particles
larger than $\sim10\micron$ can propagate for tens of parsecs through
the interstellar medium, opening up the possibility that ground based
radar systems can detect AGB stars, young stellar objects such as T
Tauri stars, and debris disks around Vega-like stars. We provide
analytical and numerical estimates of the ejection velocity in the
case of a debris disk interacting with a Jupiter mass planet. We
provide rough estimates of the flux of large micrometeoroids from all
three classes of sources. Current radar systems are unlikely to detect
significant numbers of meteors from debris disks such as Beta
Pictoris. However, we suggest improvements to  radar
systems that should allow for the detection of multiple examples of
all three classes.
\end{abstract}


\section{\label{sec:intro} INTRODUCTION}

In astronomy, as in everyday life, most of our information comes to us
in the form of electromagnetic radiation.  Some astronomical systems
also emit solid particles, which could yield valuable information if
detected.  The dust detectors on the {\it Ulysses} and {\it Galileo}
spacecraft have yielded preliminary information on the flux of
interstellar grains passing through the solar system (Frisch et
al.~1999; Landgraf et al.~2000).  Due to the limited area of the
collecting surfaces, of order $200\cm^2$, these authors were unable to
report fluxes for grains with masses $\gtsim 10^{-10} \g$.  These
low-mass grains cannot be traced to their point of origin, since
interstellar influences (e.g., gas drag, radiation pressure, magnetic
fields) rapidly adjust their velocities.

Particle fluxes probably decline dramatically as the grain mass increases,
necessitating dust detectors with much larger collecting areas.
Meteor-tracking radar facilities can be used to detect grains
originating beyond the solar system, if the grain's initial velocity can 
be inferred with sufficient accuracy.  The effective collecting area in 
this case could be $\gtsim 10^4 \, {\rm km}^2$.

Recently, Baggaley and coworkers reported the detection of extra-solar
grains using the Advanced Meteor Orbit Radar, or AMOR (Baggaley et
al.~1994; Baggaley 2000).  Baggaley (2000) finds a ``discrete source'' with
an angular diameter of $\sim 30\arcdeg$, which he tentatively
associates with $\beta$ Pictoris. Inspection of his Fig. (2) suggests
the presence of a broad, band-like feature as well.  Baggaley does not
provide any estimate of the particle fluxes.

Our goal here is to identify likely sources of extra-solar grains and
estimate expected fluxes at Earth, as a function of the grain size.
We consider three types of objects that could potentially yield
significant fluxes: young main sequence stars, asymptotic giant branch
stars (AGB stars), and young stellar objects (YSOs).

For a given source of interest, the flux at Earth depends 
on the following three factors:  1.  The ``dust luminosity'' of the source, 
i.e., the rate at which grains are emitted.  We also need
to know whether or not the emission is isotropic.  If not, then we need 
to know the orientation of the source.  2.  The distance between the source
(at the time when the grains were ejected) and the Sun (now).  3.  The 
probability that the grain survives the trip and is not deflected on its way.
Although deflected grains can be detected at Earth and will contribute to
the general dust background, they do not reveal their source.  

Define the ``specific dust luminosity'' $L_{v,a}(t, \vej, a)$ by
\be
L(t) = \int d\vej \, da \, L_{v,a}(t, \vej, a)~~~,
\ee
where $t$ is the age of the source,
$L(t)$ is the dust luminosity, $\vej$ is the speed at
which grains are ejected, and $a$ is the grain radius.\footnote{Throughout 
this paper we make the simplifying assumption that the grains are spheres.} 
In order to calculate the dust flux at Earth, it is most convenient to 
consider the reference frame in which the source is stationary.  We assume
that, in this frame, large grains simply travel radially outward; i.e., 
we ignore the Galactic potential and gravitational interactions with 
individual stars, as well as any other influences that could deflect the 
grains (see \S \ref{sec:magnetic_deflection}).  
The number density of grains at distance $d$ from the source 
is given by $n(t, d) = \int d\vej \, da \, n_{v,a}(t, \vej, a, d)$,
where
\be 
n_{v,a}(t, \vej, a, d) = \frac{L_{v,a}(\vej, a, t_{\rm ej} = t-d/\vej)}
{4 \pi d^2 \vej} f_{\rm beam} \, f_{\rm survive}(\vej, a, d)~~~;
\ee
$t_{\rm ej}$ is the age of the source when the grains are ejected.
The factor $f_{\rm beam}$ accounts for anisotropic emission from the 
source (we assume that it is independent of $\vej$ and $a$) and 
$f_{\rm survive}(\vej, a, d)$ is the fraction of the grains that survive the 
trip out to distance $d$ without being destroyed or significantly 
deflected.  The particle flux at Earth at the present time is given by
$F(t) = \int d\vej \, da \, F_{v,a}(t, \vej, a, d)$, where
\be
\label{eq:specific_flux}
F_{v,a}(t, \vej, a, d) = n_{v,a}(t, \vej, a, d) v_{d, \sun} = 
\frac{L_{v,a}(\vej, a, t_{\rm ej} = t - d/\vej)}{4 \pi d^2} \frac{v_{d, \sun}}
{\vej} f_{\rm beam} \, f_{\rm survive}(\vej, a, d)~~~.
\ee
The velocity of the dust grain with respect to the Sun is given by
\be
\label{eq:v_dust_Sun}
\vec{v}_{d, \sun} = \vec{v}_{\ast, \odot} - \vej \hat{r}_{\ast, \sun}~~~,
\ee
where $\vec{v}_{\ast, \sun}$ is the velocity of the source with respect
to the Sun and $\hat{r}_{\ast, \sun}$ is the unit vector pointing from the 
Sun to the source.

In \S \ref{sec:detectable_fluxes}, we estimate the limiting particle
flux that could potentially be detected with radar facilities.   
In \S \ref{sec:survive}, we consider the mechanisms that can prevent 
a grain from reaching us and estimate the (size-dependent) distance that a
grain can travel and still reveal its source.  We estimate observable
dust fluxes from young main sequence stars, AGB stars, and YSOs in 
\S \S \ref{sec:Vega-like}, \ref{sec:AGB}, and \ref{sec:YSOs},
respectively.  We discuss our results in \S \ref{sec:Discussion}, and
summarize our conclusions in \S \ref{sec:conclusions}.

\section{\label{sec:detectable_fluxes} DETECTABLE FLUXES}
\subsection{Satellite Measured Fluxes}
There have been several claimed detections of extrasolar meteors, both
from satellite detectors and ground based radar. The most convincing
detections are those of the Ulysses and Galileo satellites. Frisch et
al. (1999) and Landgraf et al. (2000) show that the flux of
$m\approx6\times10^{-13}\g$ extrasolar particles is given by
$mf_m\approx10^{-9}\cm^{-2}\s^{-1}$. Here $f_mdm$ is the flux of
particles with masses between $m$ and $m+d m$.\footnote{ The cited
authors use the cumulative flux in a logarithmic mass interval,
$f_{\log_{10} m}d(\log m)=f_mdm$. The two fluxes are related by
$mf_m=f_{\log_{10} m}/\ln10$. } At the low mass end of the mass
distribution (below $m\approx6\times10^{-13}\g$) the particles are subject to
strong perturbations from the solar wind and its associated magnetic
field, so that the measured flux is not representative of the flux of
small particles outside the heliosphere. Above $m\approx 6\times10^{-13}\g$,
they find $mf_m\propto m^{-1.1}$; there are
roughly equal masses of particles in every logarithmic mass bin.

For purposes of extrapolation we will occasionally assume that the
cumulative flux of
interstellar particles of mass $m$ is 
\be \label{eq:observed flux}
f_{\log_{10} m}\approx3\times10^{-9}
\left({6\times10^{-13}\g\over m}\right)^{1.1}
\cm^{-2}\s^{-1},
\ee 
for $m>6\times10^{-13}\g$. 

We employ two other differential fluxes. The first is $f_a$, where
$f_ada$ is the flux of particles with radii between $a$ and $a+da$;
$f_a$ has units $\cm^{-2}\s^{-1}\cm^{-1}$. In many cases one assumes
that $f_a$ follows a power law distribution, $f_a\propto
a^{-\gamma}$. This corresponds to $f_m\propto m^{-\alpha}$, with
$\alpha=(\gamma+2)/3$. We also use the differential mass flux,
$m^2f_m$, with units $\g\cm^{-2}\s^{-1}$.

In the asteroid literature one encounters the Dohnanyi (1969) law,
$f_m\propto m^{-\alpha}$, with $\alpha=11/6$, corresponding to
$\gamma=7/2$. The Dohnany law represents a singular, steady state
solution to a set of equations describing a closed system of colliding
bodies. Asteroids do not follow this relation\cite{Ivezic}. The
same scaling appears in the interstellar dust literature, in the MRN
dust model (Mathis, Rumpl, \& Nordsieck 1977). In these models the
mass in large particles diverges in proportion to $\sqrt{a}$, or as
$m^{1/6}$. The observed scaling (eqn. \ref{eq:observed flux}) is
$\alpha\approx2.1$, slightly larger than the Dohnanyi value, so that
the mass in large particles is finite. We note that particles in
debris disks or in the outflows of AGB stars need not be distributed
according to either law; in fact the findings below suggest that they are not
well described by a power law with $\gamma=7/2$.

\subsection{Radar  Fluxes}
Ground based radar at Arecibo \citet{2002ApJ...567..323M} and in New
Zealand (AMOR) \citet{TBS96, Bag00} have also reported
detections of extrasolar meteors. When a meteoroid enters the Earth's
atmosphere, air molecules ablate and ionize material from the
meteoroid. The free electrons created by this ablation reflect radio
waves, a fact exploited by radar aficionados.

The size of individual meteoroids detected by radar systems can be
inferred by three different methods; by the radar power reflected from
a Fresnel zone of the meteor trail (\S \ref{sec:Fresnel}), by the
power reflected from the meteor head (\S \ref{sec:head}), or by the
deceleration of the meteoroid as it ablates in the Earth's atmosphere
(also in \S \ref{sec:head}).

The first two size estimates rely on the power of the reflected radar
signal. For radar micrometeors the relevant reflection is
coherent. The wavelengths employed are comparable to or larger than
the initial width $r_0$ of the ionization trail, so the radar is
sensitive to the electron line density $q$ rather than the space
density. The reflected power depends on the transmitted power, the
distance between the radar and the meteoroid, the mass and velocity of
the meteoroid, and the (uncertain) ionization efficiency. Here we
quantify this relation. A good introduction to the subject of radar
meteors is given by \citet{McKinley61}.

\subsubsection{Meteoroid properties\label{sec:Meteor properties}}
The radii of radar meteoroids range from $a\sim3\mum$ to
$\sim40\mum$. For purposes of illustration, we consider meteoroids
with radius $a=10\mum$, density $\rho=3\g\cm^{-3}$, and  made of
silicates such as Forsterite, Mg$_2$SiO$_4$. Our default meteoroid has a
mean molecular weight of $140$ grams per mole, and a vaporization
temperature of order 2000K. We assume that the energy per bond is
$\sim2$ eV. This is equivalent to a heat of ablation of
$\zeta\sim10^{11}$ ergs per gram. 

The mass of our default meteoroid is $m=1.3\times10^{-8}\g$. The mean
atomic weight is $\mu\approx20m_p$, where $m_p=1.67\times10^{-24}\g$
is the mass of a proton, while the total number of atoms in the meteoroid
is $N=m/\mu\approx4\times10^{14}$. The total binding energy of the
meteoroid is 
\be \label{binding}
E_{B}\equiv\zeta m\approx 1300 \left({a\over 10\mic}\right)^3
\left({2\, {\rm eV}\over{\rm  bond}}\right)
\left({\rho\over 3\g\cm^{-3}}\right){\, \rm ergs}.
\ee 

A typical meteor is highly supersonic, with velocity $v\sim40\km/\s$
at the top of the atmosphere, so the kinetic energy of the meteoroid is
\be 
KE\approx 10^5\left({a\over 10\mic}\right)^3
\left({v\over 40\km\s^{-1}}\right)^2
\left({\rho\over 3\g\cm^{-3}}\right){\,\rm ergs},
\ee 
much larger than the binding energy. 

\subsubsection{The ionization trail}

As the meteoroid passes through the atmosphere, it collides with air
molecules and is ablated. Because the kinetic energy per meteoroid atom
is in excess of $100$ eV, while the ionization potential is of order
10eV (depending on the atomic species), collisions between ablated
atoms and air molecules may ionize either or both particles.  The
number of ions produced by each meteoroid atom is denoted by $\beta$ in
the meteor literature. It is electrons from these ionized atoms that
reflect the radar signal.

If the mass loss rate of a meteoroid traveling at velocity $v$ is
$dm/dt$, the number of ions produced per centimeter along the flight
path is
\be \label{q}
q=-{\beta\over \mu v}{dm\over dt}.
\ee 
The rate at which the meteoroid ablates depends on the ambient air
density, which increases exponentially as the meteoroid descends
through the atmosphere. In other words, the mass loss rate increases
rapidly with time. The pressure scale height $H_p$ of the atmosphere
is roughly constant, since $H_p=kT/(\mu_{air} g)$ and the air
temperature $T\approx200{\rm \,K}$. The line density $q$ reaches a
maximum just before the meteor vanishes, with most of the matter being
deposited in the last scale height of its path. Taking $-vdt=H_p$, we
find
\be \label{q approx}
q={3\over2}{\beta m\over \mu H_p},
\ee 
where the  factor of $3/2$ comes from the more accurate calculation
outlined in the appendix.

The value of $\beta$ depends strongly on velocity $v$, but the range
of $v$ for typical extrasolar meteors at Arecibo or AMOR is
$20-60\km\s^{-1}$, with most near $40\km\s^{-1}$. Jones (1997) and
Jones \& Halliday (2001) examine $\beta$ using a combination of
laboratory experiments and observational data. For iron atoms they
find $\beta\approx0.6(v/40\km\s^{-1})^{3.12}$ (with
$v<60\km\s^{-1}$). For more realistic compositions they expect $\beta$
to be a factor of $5$ smaller. For a meteoroid with mean atomic weight
$\mu=20m_p$ (as for Forsterite),

\be 
q\approx6\times10^7\left({\beta\over 0.1}\right)
\left({a\over10\mic}\right)^3\left({\rho\over3\g\cm^{-3}}\right)
\left({6\km\over H_p}\right)\cm^{-1}.
\ee 

The height at which the meteoroid ablates can be estimated by finding the
height at which the binding energy of the meteoroid decreases most
rapidly. The rate of change of the binding energy is
\be \label{ablation height} 
\zeta{dm\over dt}= -\Lambda \cdot\pi a^2\cdot{1\over2}\rho_av^3,
\ee 
where $\rho_a$ is the mass density of the atmosphere at the height of
the meteoroid and $\Lambda$ is the fraction of the kinetic energy flux that goes
toward ablating the meteor. Taking the derivative of this expression
with respect to time and setting the result equal to zero yields the
relation 
\be 
-{2\over3m}{dm\over dt}={1\over\rho_a}{d\rho_a\over dt}+{3\over
v}{dv\over dt}.
\ee 
We show in the appendix that the last term is typically smaller than
either of the other two, or at best comparable, so we ignore it. Using
equation (\ref{ablation height}), and setting $d\rho_a/dt=\rho_a
v/H_p$, we find that the meteoroid ablates when the atmospheric density is
\be 
\rho_a^*=\rho {4\over\Lambda}\left({a\over H_p}\right)
\left({\zeta\over v^2}\right)
\ee
Using typical values, we find
\be \label{eqn: density}
\rho_a^*\approx1.3\times10^{-9}\left({\rho\over 3\g\cm^{-3}}\right)
\left({10^{-1}\over \Lambda}\right)\left({a\over 10\mum}\right)
\left({\zeta\over10^{11}\cm^2\s^{-2}}\right)\left({40\km\s^{-1}\over  v}\right)^2
\left({6\km\over H_p}\right)
\ee 

Using an MSIS model atmosphere\footnote{Available at
http://nssdc.gsfc.nasa.gov/space/model/models/msis.html} we find that
this density occurs at a height of $96\km$, comparable to
the observed height at both AMOR and Arecibo.  

On the other hand, radar data from Jodrell Bank and from Ottawa show
that radar meteoroids detected by those systems also ablate at heights
near $95\km$ (McKinley 1961), despite the fact that they have line
densities $q\approx10^{14}\cm^{-1}$, six orders of magnitude larger than the
Arecibo or AMOR meteoroids traveling at the same velocity. These larger
meteoroids are still much smaller than the mean free path, so that
aerodynamic effects seem unlikely to explain the difference;
nevertheless, something must differ between the two classes of
objects, else the more massive Jodrell Bank meteoroids would be ablated
five scale heights below the less massive meteoroids discussed here. One
possibility that has been suggested is that the more massive meteoroids
(those seen at Jodrell Bank and Ottawa) fragment into smaller pieces,
which are then rapidly ablated. The observed heights of the Arecibo
and AMOR events suggest that for micrometeoroids that do not fragment
$\Lambda\approx 0.1$ and $\zeta\approx10^{11}\cm^2\s^{-2}$.

We stated above that the initial width $r_0$ of the ionization trail
was comparable to or smaller than the radar wavelength. We estimate
$r_0$ as follows. The cross-section of an atom is roughly $\sigma=\pi
d^2$, with $d\approx2\times10^{-8}\cm$; at 95km the MSIS atmosphere has
$T=171K$, and a mean free path of $l\approx37\cm$. However, the
typical meteoroid atom has an atomic mass about twice that of a nitrogen
atom. In a collision between a meteoroid atom and a nitrogen molecule,
the momentum will not be shared equally between the two N atoms; the
binding energy of the $N_2$ is much less than the kinetic energy of
the meteoroid atom. Neglecting the momentum carried away by the slower of
the two rebounding N atoms, we find that the final velocity of the
meteoroid atom is about $2/3$ of its initial velocity for a head on
collision. After $n$ such collisions, the velocity is
$v_n=(2/3)^nv_0$. Setting $v_n$ equal to twice the thermal velocity at
$95\km$, we find that the meteoroid atom is thermalized after $n\sim10$
collisions. Thus $r_0\approx\sqrt{n}\,l$, or about $100\cm$. A strict
upper limit to the thermalization time is $\tau_{th}=l/v$, about
$10\mu\s$, much shorter than the time between radar pulses. Using
different arguments, \cite{Manning} and \cite{Bronshten} obtain
similar results.  After this initial rapid diffusion, ordinary thermal
diffusion sets in and the trail radius slowly increases.

The initial electron density is 
\be 
n_e=q/\pi
r_0^2\approx2\times10^3\cm^{-3}\left({\beta\over 0.1}\right)
\left({a\over10\mic}\right)^3\left({6\km\over H_p}\right)
\left({\rho\over 3 \g\cm^{-3}}\right)\cm^{-3}
\ee 

The plasma frequency $\nu_p \approx 0.39 {\rm\,MHz}$ is
well below the frequency employed in the radar systems, so the radar
beam will penetrate through the trail, ensuring that all the electrons
will reflect. The average distance between electrons is
$n^{-1/3}\approx 0.1\cm$, much less than the wavelength of either AMOR
or Arecibo, so the reflected emission will be coherent in both
systems.

The thermal diffusion mentioned above eventually causes the trail
radius to exceed the wavelength of the radar. This diminishes the
visibility of the trail, particularly for trails formed at great
heights, or for short wavelength radar. A recent discussion of this
'height ceiling' can be found in Campbell-Brown \& Jones (2003).

\subsubsection{Minimum detectable meteoroid size at AMOR\label{sec:Fresnel}}
We proceed to calculate the power collected by the radar receiver at
AMOR. The relevant properties of the AMOR radar (as
well as those for the Arecibo setup) are listed in Table 1.  An
electron a distance $R$ from the radar sees a flux $P_TG_T/4\pi R^2$,
assuming the radar transmits a power $P_T$ and has an antenna gain
$G_T$. The gain is defined as $4\pi$ steradians (the solid angle into
which an isotropic emitter radiates) divided by the solid angle of the
radar beam; the latter is roughly $(\lambda/D)^2$, where $D$ is the
radius or length of the antenna. Hence the effective area of the
antenna is related to the gain by $A_{eff}=G_T\lambda^2/4\pi$. The
electron scatters the incident wave, emitting a power
$(3/2)P_TG_T\sigma_T/4\pi R^2$ back toward the radar; the factor $3/2$
arises because Thompson scattering is that of a Hertzian dipole rather
than isotropic.  The receiving antenna (which in the case of Arecibo
is the same as the transmitting antenna) captures a fraction
$A_{eff}/4\pi R^2$ of this scattered power. The power received at the
antenna from a single electron is
\be \label{power} 
\Delta P_e = P_T\left({3G_TG_R\over 128\pi^3 }\right)
\left({\sigma_T\lambda^2\over R^4}\right).
\ee 
We have allowed for the possibility that the gain for the receiver and
transmitter are different, as is indeed the case at AMOR.

The amplitude of the
electric field at the receiver is
\be \label{amplitude} 
\Delta A_e=\sqrt{2r\Delta P_e},
\ee 
where $r$ is the impedance of the receiver.

The instantaneous signal amplitude received from a meteor trail is found by summing
over all the electrons along the track:
\be 
A(t)=\int_{s_1}^{s_2}\left(2r\Delta P_e\right)^{1/2}q(s)\sin(\omega t-2kR)ds,
\ee 
where $\omega$ is the angular frequency of the radar, and
$k=2\pi/\lambda$. The total distance from the radar to the trail and
back is $2R$, whence the factor $2$ in the argument of the sine. We
measure $s$ along the trail starting from the point $X$ on the trail
at which $q$ reaches its maximum. The distance between the radar and
point $X$ is denoted by $R_0$, while the angle between the line of
sight and the meteor trail at $X$ is denoted by $\pi/2+\theta$.

Near $X$,
\be 
R\approx R_0\left[1+{s\over R_0}\theta+{1\over2}\left(s\over R_0\right)^2\right],
\ee 
where we assume $\theta<<1$. Define $\chi=\omega t-2kR_0$, and
$z=2s/\sqrt{R_0\lambda}$. 
Then
\be 
A(t)=\sqrt{2r\Delta P_e}q{\sqrt{R_0\lambda}\over 2}
\int_{z_1}^{z_2}\sin\left[\chi-2\pi\sqrt{R_0\over\lambda}\theta z-
{\pi z^2\over2}\right]dz
\ee 
When $\theta<<\sqrt{\lambda/R_0}$ the integral reduces to a Fresnel
integral, which is an oscillating quantity of order unity. The peak
power at AMOR is then
\be \label{AMOR power final}
P_R\approx P_T {3G_TG_R\over256\pi^3}
\left({\lambda\over R_0}\right)^3q^2\sigma_T.
\ee 
When $\theta\ga\sqrt{\lambda/R_0}$, the integral cuts off at $z\la 1/4$.
In other words, the AMOR
detector is sensitive only to micrometeor trails within an angle
$\theta\approx\sqrt{\lambda/ R_0}$ of the perpendicular to the
line of sight.

The minimum detectable line density for AMOR is
\be \label{q min}
q_{min}={1\over \sqrt{\sigma_T}}\left(P_R\over P_T\right)^{1/2}
\left(256\pi^3\over 3G_TG_R\right)^{1/2}\left({R_0\over
\lambda}\right)^{3/2}.
\ee 
The minimum detectable power at AMOR is $P_n=1.6\times 10^{-13}$
Watts, the transmitted power is $P_T=100$kW, while the gains are
$G_T=430$ and $G_R=130$. The typical range is $R_0\approx200\km$,
leading to
\be 
q_{min}\approx7\times10^8\cm^{-1}.
\ee 
Using equation (\ref{q approx}), the minimum detectable mass is
\be 
m_{min}\approx2\times10^{-7}\g,
\ee 
corresponding to a minimum radius of
\be 
a_{min}\approx25\left({3\g\cm^{-3}\over\rho}\right)^{1/3}
\left({0.1\over\beta}\right)^{1/3}
\left({R_0\over200\km}\right)^{2/3} \mu m.
\ee 

We are now in a position to estimate the collecting area $A_{coll}$ of
the AMOR detector (as a meteor detector, not as a radar). The
collecting area is the product of the sensitivity weighted geometric
area $A_G$ imaged by the radar and the fraction of solid angle
$\sqrt{\lambda/R_0}$ to which it is sensitive;
\be 
A_{coll}\approx \sqrt{\lambda\over R_0}A_G.
\ee 
We define the sensitivity weighted geometric area as
\be 
A_G\equiv\int {F(m)\over  F(m(R_{min}))}R\,dR\,d\phi .
\ee 
Here $F(m)$ is the flux of particles of the minimum detectable mass
$m=m_{min}(R)$; the latter is a function of $R$, the distance between
the radar and the meteor trail. We assume that $m_{min}\propto
q_{min}\propto R^{3/2}$, where the last scaling comes from equation
(\ref{q min}), and that $F(m)\sim m^{-1}$ as an analytically simpler
version of (\ref{eq:observed flux}). With these assumptions,
\be 
A_G\approx 2R_{min}^2\Delta\phi{\left[\sqrt{R_{max}/R_{min}}-1\right]}
\ee 
The AMOR beam has a width of $\Delta \phi \approx3^\circ$, and
extends from $12^\circ$ to $30^\circ$ above the horizon. We further
assume that all the meteors are detected at heights of
$\sim90\km$. This implies that the lower and upper limits to the
integration are $R_{min}=165\km$ and $R_{max}=430\km$. We
find $A_G\approx10^{13}\cm$. The collecting
area of the radar system is then 
\be 
A_{col}\approx 8\km^2.
\ee 

\begin{figure}
\epsscale{1.00}
\plotone{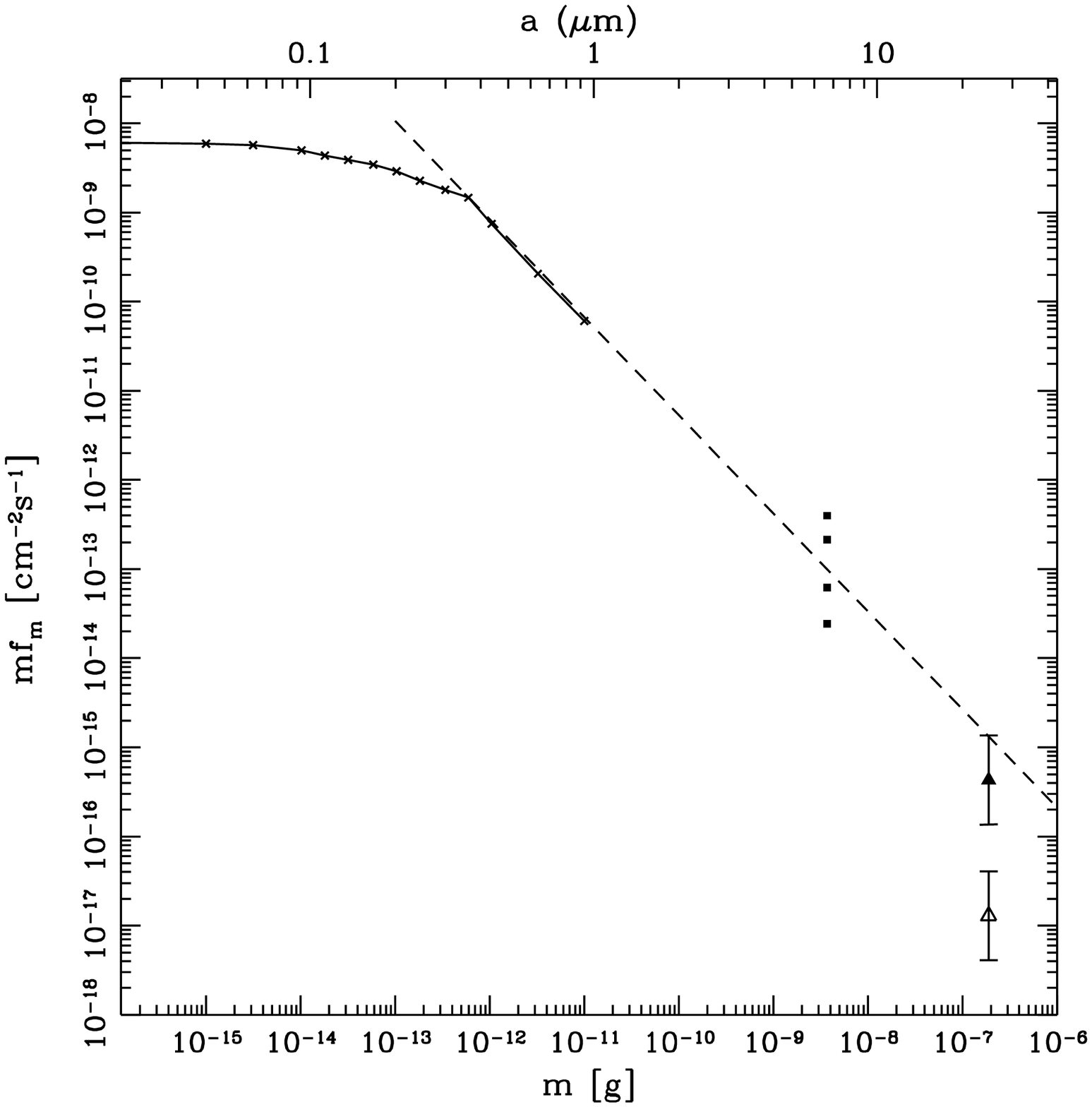}
\caption{ Measured cumulative fluxes of extrasolar meteoroids. The data
are from the Ulysses and Galileo satellites (crosses joined by a solid
line), the Arecibo radar (solid squares) and the AMOR radar (solid
triangle). The open triangle represent the ``point source'' seen by
the AMOR radar. The dashed line represents the $mf_m\propto m^{-1.1}$
scaling first noted by Landgraf et al. (2000) and given in equation
(\ref{eq:observed flux}). The upper axis gives the particle radius $a$
in microns, assuming that the meteoroids have a density of $3\g\cm^{-3}$.
\label{fig:flux}
}
\end{figure}

From Fig. 1b in Baggaley (2000), we estimate that in 4 years of
continuous operation  AMOR detected  $\sim10^4$ extrasolar meteors, so
we estimate the flux as
\be 
mf_m\left(m=2\times10^{-7}\g\right)\approx 4\times10^{-16}
\cm^{-2}\s^{-1}.
\ee 
From the same paper, we estimate that the number of extrasolar meteors
detected from Baggaley's ``point source'' to be about about $200$,
leading to a flux $50$ times smaller. Both fluxes are plotted in
Fig. (\ref{fig:flux}), along with flux estimates from Ulysses and
Galileo, and from Arecibo.

\begin{figure}
\epsscale{1.00}
\plotone{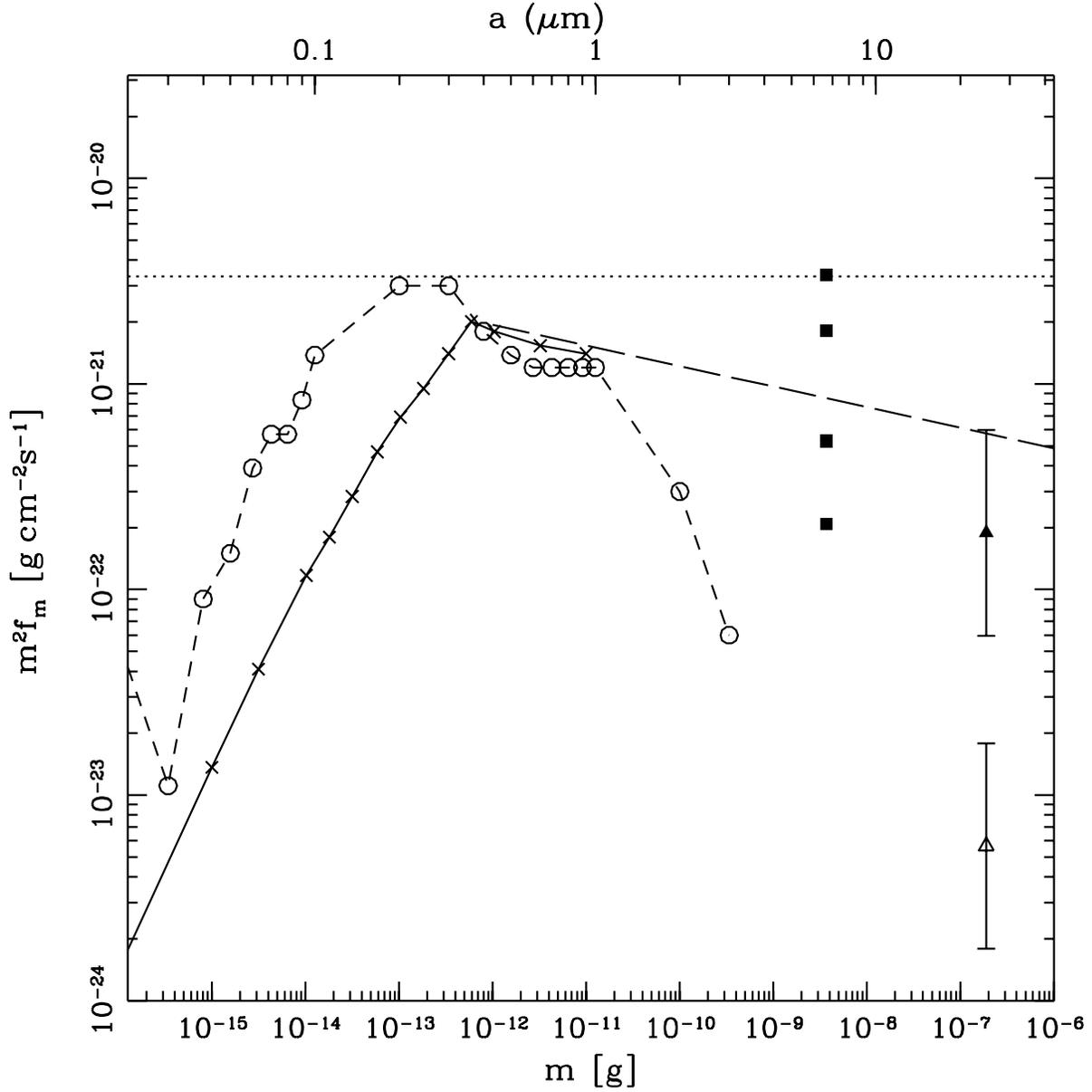}
\caption{ Measured differential mass fluxes, $m\cdot mf_m$, with units
  of $\g\cm^{-3}\s^{-1}$, of extrasolar micrometeoroids. The
  long-dashed line is the Landgraf et al. (2000) scaling, while the
  dotted line is the limit found by assuming that half of all the
  metals in the local ISM (with $n=0.1\cm^{-3}$) are in the form of
  micrometeoroids. The solid squares are the fluxes reported by the
  Arecibo radar. The filled triangle is the flux estimated from the
  spatially distributed flux in Fig. 2 of Baggaley (2000). The open
  triangle is the flux in Baggaley's ``point source''. The horizontal
  dotted line is the upper limit on the flux given by assuming that
  half of all the heavy elements are in dust grains. The dashed line
  joining open circles is the flux corresponding to the ISM dust size
  distribution found by Kim and Martin (1995), normalized to the
  dotted line. This figure suggests that the bulk of the mass in dust
  consists of particles with masses of order $10^{-13}\g$, or sizes of
  order two tenths of a micron.
\label{fig:mass flux}
}
\end{figure}

Fig. (\ref{fig:mass flux}) shows the same data as a differential mass
flux. In this plot the bulk of the mass flux is represented by the
highest point, making it easy to see that most of the mass of
interstellar meteoroids reaching the inner solar system comes in the form
of $\sim 0.3\micron$ objects. On the same figure we have plotted the
size distribution found by Kim \& Martin (1995), normalized to a
hydrogen number density of $0.1$ and assuming that half the metals are
in the form of dust grains. 

\subsubsection{Minimum detectable meteoroid size at Arecibo \label{sec:head}}
The minimum detectable meteoroid size for Arecibo can be calculated in a
similar manner, starting from equations (\ref{power}) and
(\ref{amplitude}). Arecibo uses a much shorter wavelength than AMOR,
$70\cm$ rather than $1145\cm$. Arecibo also has a much narrower beam
than AMOR, $\sim500\m$ versus $\sim10\km$. It is thus insensitive to trails
traveling across the beam, in contrast to AMOR. Due to the fact that
trails observed perpendicular to the line of sight have long coherence
lengths, AMOR is sensitive to fairly small meteoroids despite the rather
modest power and antenna gains employed. Arecibo, by contrast, is
sensitive primarily to vertical trails, which have short coherence
lengths (a quarter of a wavelength, less than $20\cm$). This is
compensated for by the much larger gains, and by the larger
transmitted power of the Arecibo radar.

The short wavelength employed makes the Arecibo radar sensitive to
diffusion of electrons across the meteoroid trail. The Arecibo radar beam
is roughly vertical, with a half opening angle of $1/6^\circ$, so at a
height of $100\km$ it is about $500$ meters wide. Typical Arecibo
meteor trails are a few kilometers long, so they are within a tenth of
a radian or so of the vertical. Thus the ion trail left by an Arecibo
meteoroid diffuses primarily horizontally, since the density along the
trail varies on kilometer scales. As shown above, the initial trail
width is of order $100\cm$, but this width increases over time due to
diffusion. The diffusion coefficient $D\sim c_sl\sim3\times
10^5\cm^2\s^{-1}$, where $c_s\approx3\times10^4\cm\s^{-1}$ is the
sound speed, and $l\approx10\cm$ is the mean free path of air
molecules. In the horizontal direction the Fresnel length is
$L_F\equiv\sqrt{R_0\lambda}\approx 3\times10^4\cm$, much larger than
the initial beam width $r_0\approx100\cm$. The time to diffuse
horizontally over a Fresnel zone is $L_F^2/D\approx3000\s$, so
diffusion is irrelevant for a vertical trail. However, for a trail
$0.1$ radians off the vertical, the time to diffuse $\lambda/4$ in the
vertical direction is about 13 ms. This will limit the time the beam
remains visible to the radar at a given height to about 13 ms for a
typical meteor. This is consistent with the trail lengths observed by
Mathews et al. (1997).

We have already noted that the meteor trail is much narrower than the
Fresnel length of the Arecibo radar; then all the
electrons in a patch of length of order $\lambda/2$ along the trail
radiate coherently. Suppose that $N_{ch}$ is the number of coherently
emitting electrons in such a patch. The
amplitude of the electric field at the receiver from a single patch is
\be \label{Arecibo Amplitude}
A\approx\sqrt{2r\Delta P_e}N_{ch}.
\ee 
The vertical resolution of the Arecibo antenna is $L\approx 150\m$, so
that there are $N_p\sim L/\lambda\approx200$ 
patches radiating in a resolution length. The signal observed at
Arecibo varies on a length scale much shorter than the atmospheric
scale height, indicating that the mass loss rate fluctuates
rapidly. We assume that these fluctuations occur on scales smaller
than $L$, so that the electric field from each patch adds
incoherently.
The power received by the detector is 
\be \label{Arecibo power} 
P_{\rm Arecibo}=P_T\left({3G_TG_R\over128\pi^3}\right)
\sqrt{L\over\lambda}\left({\sigma_T\lambda^2\over R_0^4}\right)N_{ch}^2.
\ee 
Since the coherence length along the (assumed vertical) trail is of
order $\lambda$, the number of coherently scattering electrons 
is simply related to the line density,
\be 
N_{ch}\approx q\cdot\lambda/2.
\ee 
The expression for the power becomes
\be \label{Arecibo power final}
P=P_T\left(3G_TG_R\over512\pi^3\right)
\sqrt{L\over\lambda}\left({\lambda^4\over R_0^4}\right)
q^2\sigma_T.
\ee 

For a given transmitted power, wavelength and antenna gain, this is smaller than
the equivalent expression (\ref{AMOR power final}) by the ratio
\be 
{\sqrt{\lambda L}\over R_0}\approx10^{-4},
\ee 
using values for Arecibo. Worse, the power scales as $\lambda^3$ (or
as $\lambda^{3.5}$ for Arecibo), so the $\sim15$ times shorter
wavelength employed by Arecibo reduces the received power by another
factor of $\sim3000$. However, these two factors are more than
compensated for by the much larger gains of the Arecibo antenna. The
net effect is that despite its higher power and much larger gain, the
Arecibo radar can detect meteoroids that are a factor of seven to ten
smaller in radius than the AMOR setup, but no smaller.

Define the signal to noise ratio $SNR\equiv P/P_{n}$, where
$P_n$ is the noise power for the Arecibo receiver.  Using this definition, and
solving for $q$,
\be \label{eqn:Arecibo q}
q={1\over \sqrt{\sigma_T}}\left({P_n\over P_T}\right)^{1/2}
\left({512\pi^3\over 3G_TG_R}\right)^{1/2}
\left({\lambda\over L}\right)^{1/4}
\left({R_0\over \lambda}\right)^2
SNR^{1/2}
\ee 

Using the values in Table 1, we find
$q\approx10^7\cm^{-1}$. Using equation (\ref{q approx}) we find
\be 
m\approx3\times10^{-9}\left({0.1\over\beta}\right)
\left({R_0\over100\km}\right)^2 \left({150\m\over L}\right)^{1/4}SNR^{1/2}g.
\ee 
The corresponding meteoroid radius is
\be \label{radar size}
a \approx6\left({3\g\cm^{-3}\over\rho}\right)^{1/3}
\left({0.1\over\beta}\right)^{1/3}
\left({R_0\over100\km}\right)^{2/3} 
\left({150\m\over L}\right)^{1/12}
SNR^{1/6}\mu m.
\ee 

A firm lower limit for the size of the meteoroids seen at Arecibo is
found by noting that the energy per meteoroid atom is $\sim100$ eV. Since
the ionization potential of either atoms or molecules is typically of
order $10$ eV, a single meteoroid atom can free at most $10$
electrons. Taking $\rho=3\g/\cm^3$ and $\beta=10$, we find
$a_{min}\approx1.3\micron$.

From equation (\ref{Arecibo power final}), and noting that
$G_R\lambda^2=const.$, we see that, as long as $\lambda$ is larger
than the initial radius $r_0$ of the trail, $P\propto
\lambda^1.5$. Some Arecibo meteors are seen at both 430 MHz and 50 MHz
(Meisel, private communication). This suggests that the volume
occupied by the reflecting electrons has linear dimensions smaller
than $\lambda/2\approx35\cm$. 

For example, it might be thought that the meteoroid itself captures
enough electrons to scatter the radar beam. However, this is not
possible. The number of charges required to reflect the beam is given
by equation (\ref{Arecibo power}), except that in this case there is
only one emitting region (the meteor) so the factor $\sqrt{L/\lambda}$
should be dropped. We find $N_{ch}\approx10^9$; the meteoroid is charged
to a (negative) voltage of $V=4.8\times10^5(3\micron/a)$
volts. Charging the meteoroid to such a high voltage leads to rapid
electron emission by quantum tunneling, also known as field emission
(Fowler \& Nordheim 1928). The discharge current (in statamps per
square centimeter) is given by
\be  
I={e\over2\pi h}{(\mu/\phi)^{1/2}\over\phi+\mu}E^2e^{-4\pi \phi^3/2/3E},
\ee  
where $E\approx V/a$ is the electric field
near the surface of the meteor, $\phi$ is the work function, and $\mu$
is the Fermi energy. Both of the latter have values of order $5$ eV.

We make the optimistic assumption that the meteoroid gains an electron
for every atom it collides with in the atmosphere, $I_{in}=e n^*_a v$,
also in statamps per square centimeter ($n^*_a$ is the number density
of air molecules at the height where the meteoroid ablates, see equation
\ref{eqn: density}). Setting these two currents equal, we find the
equilibrium field $E\approx3.6\times10^7\ {\rm V}\cm^{-1}$  or a
maximum voltage of $V=10^4(a/3\micron)$. The number of charges on the meteor
cannot exceed $N_{ch, max}\approx2\times10^7(a/3\micron)^2$, a factor
of $30$ too small to explain the Arecibo observations. The meteor head
signal must be due to electrons in the atmosphere around or trailing
the meteor.

One may also estimate $a$ dynamically, if one can measure the rate of
deceleration of the meteoroid due to atmospheric drag;
\be 
m{dv\over dt}=-\Gamma\rho_av^2A_m,
\ee 
where $\Gamma$ is the drag coefficient, and $A_m$ is the area of the
meteor.
The radar measures $dv/dt$, $v$, and the height of the meteor. Using a
standard atmospheric model, the density $\rho_a$ can be determined,
For micrometeoroids, which have a size less than the mean free path for
collisions, $\Gamma=1$, so the ratio $m/A_m$ can be determined
directly from measured quantities. \cite{JMMZ} refer to this ratio as
the ballistic parameter, $BP$. We assume spherical meteoroids, so
$BP\approx 4\rho a/3$. Then a simple estimate for $a$ is
\be 
a={3\over4}{{\rm BP}\over \rho}
\ee 
\cite{JMMZ} find a range of ballistic parameters, ranging down to
$10^{-3}\g\cm^{-2}$ or slightly lower, leading to
\be \label{BP size}
a_{min}\approx2.5\left({3\g\cm^{-3}\over
\rho}\right)\left({BP\over10^{-3}\g\cm^{-2}}\right)\mu m.  
\ee 
This is a factor 2 smaller than the radius estimate based on received
radar power, or a factor of 8 smaller when considering the meteor
mass. For the meteors with the smallest BP, the size drops to $1\mu m$
or less. It is very difficult to see how the power reflected from such
a low mass meteoroid could be detected at Arecibo, suggesting that the
micrometeoroids have a density closer to $1\g\cm^{-3}$ than to
$3\g\cm^{-3}$. In fact, if we set $\rho=1$, we find $a\sim8.5\mu m$ and
$a\sim 7.5\mu m$ from the power and BP estimates, respectively. 

The fluxes observed at Arecibo are given by Meisel et al. (2002); they
are also shown on Figure 1.

\section{\label{sec:survive} PROPAGATION OF LARGE GRAINS THROUGH THE ISM}

We have seen that AMOR can only detect grains with masses 
$\gtsim 2 \times 10^{-7} \g$, corresponding to a grain radius
$a \approx 25 \micron$ for silicate dust (mass density $\rho \approx
3.5 \g \cm^{-3}$). Similarly, Arecibo can detect particles with
$a\gtsim 6\micron$. Thus, we will restrict our attention to ``large''
(by interstellar dust standards) grains, with $a \gtsim 10 \micron$.  

For such large grains in the interstellar medium (ISM), radiation pressure
can be ignored.  The force due to radiation pressure is 
$F_{\rm rad} = \pi a^2 \langle Q_{\rm pr} \rangle \Delta u_{\rm rad}$, 
where $\langle Q_{\rm pr} \rangle$ is the radiation pressure efficiency 
factor averaged over the interstellar radiation field (ISRF) and 
$\Delta u_{\rm rad}$ is the energy density in the 
anisotropic component of the ISRF.
Adopting the ISRF for the solar neighborhood (Mezger, Mathis, \&
Panagia 1982; Mathis, Mezger, \& Panagia 1983) and a 10\% anisotropy 
(Weingartner \& Draine 2001b), 
$\langle Q_{\rm pr} \rangle \approx 1$ for $a \gtsim 10 \micron$ and 
$\Delta u_{\rm rad} \approx 8.64 \times 10^{-14} \E$.  The time interval 
required for radiation pressure to change a silicate
grain's velocity by 
$1 \kms$ (assuming the anisotropy direction remains the same as the grain
moves through space) is thus given by $\Delta t \approx 5 \times 10^8 \yr
(a/10 \micron)$.  For large grains in the diffuse ISM, the forces resulting 
from the asymmetric photon-stimulated ejection of electrons and adsorbed 
atoms are even less effective than the radiation pressure (Weingartner \& 
Draine 2001b), so these forces can be neglected as well.

The following three influences can be significant for grains with 
$a \gtsim 10 \micron$:  1.  the drag force, 2.  grain destruction following
impacts with interstellar grains or high-energy gas atoms, and 3. the
magnetic force.  

We will consider the propagation of grains through two idealized phases of
the ISM, the cold and warm neutral media (CNM and WNM, respectively).  We
also consider the Local Bubble (LB), i.e., the large volume of low-density, 
ionized gas that surrounds the Sun (see, e.g., Sfeir et al.~1999).  Our 
adopted values for the gas temperature $T_{\rm gas}$, H number density $\nH$, 
and electron fraction $x_e \equiv n_e/\nH$ ($n_e$ is the electron number 
density) in these three environments are
given in Table \ref{tab:phases}.  In each case we adopt the ISRF for the 
solar neighborhood.

The drag and magnetic forces both depend on the grain's electric charge, 
which is set by a balance between the accretion of electrons from the gas 
versus photoelectric emission and proton accretion.  For large enough grains,
the rates of each of these processes are proportional to the grain area;
thus, the electric potential $\Phi$ of large grains is independent of the 
size $a$.  Using the charging algorithm from Weingartner \& Draine (2001c),
we find that silicate grains with $a \gtsim 10 \micron$ charge to 
(0.15 V, 0.74 V, 0.94 V) in the (CNM, WNM, LB).\footnote{Here, and throughout
this paper, we take silicate optical properties from Li \& Draine 2001.}  

\subsection{The Drag Force}

When a grain's motion is highly supersonic, the hydrodynamic drag (i.e., the
drag due to direct impacts of gas atoms and ions) does not depend on the gas
temperature.  In this case, the grain's speed decreases by a factor of $e$
once it has encountered its own mass in gas.  The sound speed is 
$0.81 \kms$ ($6.2 \kms$) in the CNM (WNM), assuming that the He number 
density $n_{\rm He} = 0.1 \nH$.   
We expect grain speeds as low as $\sim 10 \kms$,
so deviations from the highly supersonic limit can be expected in the
WNM.  In addition to the hydrodynamic drag, there is also the Coulomb drag,
due to long-range electric interactions between the charged grains and 
gas-phase ions.  The Coulomb drag contributes significantly to the total 
drag in the WNM for drift speeds $\ltsim 10 \kms$.
Using the approximate drag expressions from Draine \& Salpeter (1979),
we find that the actual drag force (hydrodynamic plus Coulomb)
in the WNM exceeds the hydrodynamic drag force calculated in the 
highly supersonic limit by only a factor of $\approx 2$ at a drift speed
of $10 \kms$.  For regions outside of the LB, we will adopt the highly 
supersonic limit and assume that drag
limits the distance a grain can travel to its velocity $e$-folding distance
$D_{\rm drag}$, given by
\be
D_{\rm drag} = 650 \pc \left( \frac{\rho}{3.5 \g \cm^{-3}} \right)
\left( \frac{a}{10 \micron} \right) \left( \frac{\nH}{1 \cm^{-3}} 
\right)^{-1}~~~.
\ee
Near the Sun, the average H density near the Galactic midplane is 
$\nH \approx 1 \cm^{-3}$ (Whittet 1992).

In the LB, the sound speed is $\approx 100 \kms$; thus, the highly subsonic
limit may apply.  In this case, the drag force is proportional to the grain
speed.  Using the Draine \& Salpeter (1979) drag expressions, we find that
the Coulomb drag is negligible and that
\be
D_{\rm drag} = 7.8 \, {\rm kpc} \left( \frac{\rho}{3.5 \g \cm^{-3}} \right)
\left( \frac{a}{10 \micron} \right) \left( \frac{\nH}{1 \cm^{-3}}
\right)^{-1} \left( \frac{T_{\rm gas}}{10^6 \K} \right)^{-1/2} 
\left( \frac{v_0}{10 \kms} \right)~~~,
\ee
where $v_0$ is the grain's initial speed.  Since the maximum extent of the 
LB is $\approx 250 \pc$ (Sfeir et al.~1999) and $\nH \approx 5 \times 10^{-3}
\cm^{-3}$, drag in the LB is always insignificant, whether the subsonic or 
supersonic limit applies.  

\subsection{\label{sec:magnetic_deflection} The Magnetic Force}

Charged grains spiral around magnetic field lines with a gyroradius
\be
\label{eq:gyroradius}
r_B = 17 \pc \left( \frac{\rho}{3.5 \g \cm^{-3}}\right)
\left( \frac{\Phi}{0.5 \, {\rm V}}\right)^{-1}
\left( \frac{B}{5 \mu {\rm G}}\right)^{-1}
\left( \frac{v}{10 \kms}\right)
\left( \frac{a}{10 \micron}\right)^2~~~.
\ee
Although the magnetic field strength $B$ has not been measured in the Local 
Bubble, $B \sim 5 \mu G$ just outside the LB, with the random component 
dominating the ordered component (Heiles 1998).  Thus, the magnetic force 
can significantly deflect grains with $a \approx 10 \micron$.  Since 
$r_B \propto a^2$, the importance of the magnetic deflection rapidly 
decreases with grain size.  

\subsection{Grain Destruction}

When a grain travels through the ISM, it is subjected to various agents 
of destruction, including:  1.  Sputtering, in which gas-phase ions strike 
the grain and remove grain surface atoms, which then enter the gas.  It is 
useful to distinguish between thermal sputtering (due to thermal motion 
of the gas atoms) and non-thermal sputtering (due to the motion of the 
grain with respect to the gas).  2.  Shattering and vaporization in 
grain-grain collisions.  3.  For ices, sublimation and photodesorption.

Thermal and non-thermal sputtering have been extensively discussed by 
Tielens et al.~(1994).  Applying their analysis, we find that sputtering
of grains with $a \gtsim 10 \micron$ can be neglected in all environments 
of interest.

The physics of grain-grain collisions has been treated extensively by 
Tielens et al.~(1994) and Jones et al.~(1996).  When the relative 
speed exceeds $\approx 3 \kms$ (see Table 1 in Jones et al.~1996), a crater 
forms on the larger grain (the target).  A portion of the evacuated mass 
remains on the grain as a crater lip.  The rest of the crater mass is 
removed, both as vapor and shattering fragments.  The shattering fragments 
dominate the vapor, and the largest fragment is typically somewhat larger
than the smaller impacting grain (the projectile).  When the relative 
speed exceeds $\sim 100 \kms$ (depending on the target and projectile 
materials, see Table 1 in Jones et al.~1996), the larger grain is 
entirely disrupted, and the largest shattering fragment can be a substantial
fraction of the target size.  For our purposes, the cratering regime is
generally applicable.

Jones et al.~(1996) give simple approximations for the (material-dependent)
ratio of the crater mass ($M_c$) to the projectile mass ($M_{proj}$).  
Interstellar dust is thought to be dominated by two populations:  
silicates and carbonaceous (e.g., graphite) grains.  For a head-on 
collision between a silicate target
and either a silicate or graphite projectile, the Jones et al.~(1996)
results can be approximated by
\be
\label{eq:crater_mass}
\frac{M_c}{M_{proj}} \approx 18 \left(\frac{v}{10 \kms}\right) + 29 \left(
\frac{v}{10 \kms} \right)^2~~~, \ee where $v$ is the grain-grain
relative speed.  
This empirical approximation reproduces detailed calculations, using equation 
(1) and Table 1 from Jones et al. (1996), to within 10\% when $5 \le v \le 
100 \km \s^{-1}$.    
The crater mass is
approximately 4.2 (2.2) times bigger for a graphite (ice) target.  We
will assume that equation (\ref{eq:crater_mass}) applies to all
collisions and that the entire crater mass is ejected.  This is a
conservative assumption, since the crater mass should actually be
smaller for oblique collisions, and a portion of the crater mass will
form a lip.  On the other hand, we underestimate destruction if target
grains are made of less resilient material than silicate or if the
grains are fluffy.

Since the mass in dust is $\approx 0.011$ times the mass in H (in all 
gas-phase forms) in the ISM (see, e.g., Weingartner \& Draine 2001a), the
distance a grain can travel before it loses half its mass is given by
\be
\label{eq:D_dust}
D_{\rm dest} \approx 40.2 \, {\rm kpc} \left( \frac{\rho}{3.5 \g \cm^{-3}} 
\right) \left( \frac{a}{10 \micron} \right) \left( \frac{\nH}{1 \cm^{-3}}
\right)^{-1} \left[ 18 \left(\frac{v}{10 \kms} \right) + 29 
\left(\frac{v}{10 \kms} \right)^2 \right]^{-1}~~~,
\ee
yielding $D_{\rm dest} \approx 860 \pc$ for $v = 10 \kms$ (and with other
canonical parameter values as in eq.~\ref{eq:D_dust}).

\subsection{The Survival Probability}

\begin{figure}
\epsscale{1.00}
\plotone{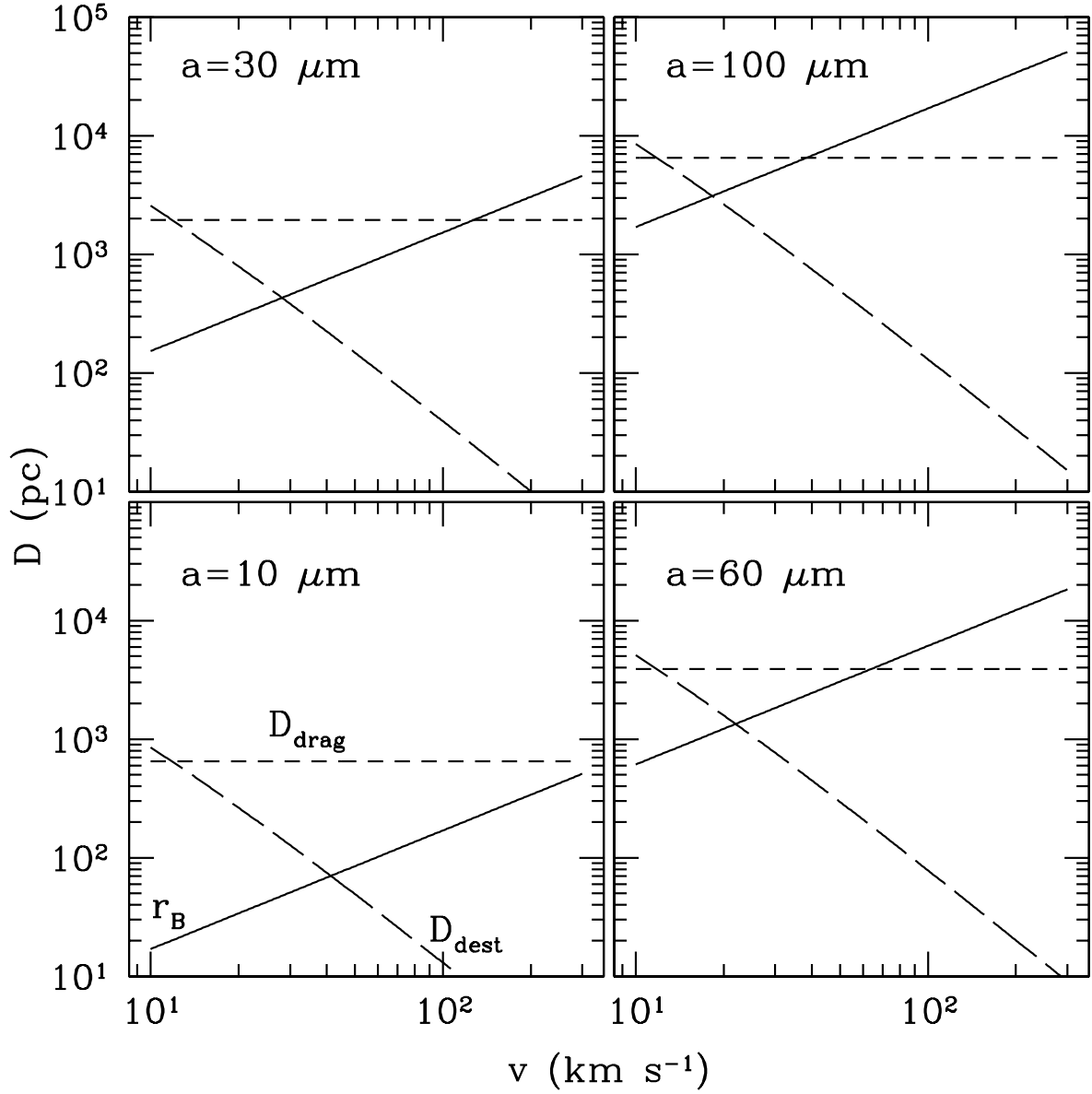}
\caption{
\label{fig:f_survive}
The drag distance $D_{\rm drag}$ (short dash), gyroradius $r_B$ (solid), and 
grain destruction distance $D_{\rm dest}$ (long dash) versus grain speed $v$ 
for four grain sizes, as indicated, and assuming supersonic grain speeds 
and $\rho = 3.5 \g \cm^{-3}$, 
$\nH = 1 \cm^{-3}$, $U=0.5 \, {\rm V}$, and $B = 5 \, \mu G$.
        }
\end{figure}

In Figure \ref{fig:f_survive}, we plot $D_{\rm drag}$, $r_B$, and 
$D_{\rm dest}$ versus $v$ for four grain sizes, covering the range of
interest.  Due to the expected dramatic decrease in flux with $a$
(e.g., see eq.~\ref{eq:flux_at_earth} below), we are not interested in 
$a \gtsim 100 \micron$.  We adopt $\rho = 3.5 \g \cm^{-3}$, 
$\nH = 1 \cm^{-3}$, $\Phi=0.5 \, {\rm V}$, and $B = 5 \, \mu G$, and assume 
supersonic grain speeds, for these plots.  For all of
the considered grain sizes, magnetic deflection dominates the other two
processes when $v = 10 \kms$, while destruction dominates when 
$v > \,$a few to several$\, \times 10 \kms$.  The drag force is never dominant.
The detailed analysis of the dynamics of a charged 
grain in a region with both ordered and random magnetic field components
is beyond the scope of this paper.  Thus, we will simply assume that
\be
f_{\rm survive} = \cases{1 &, $d < \min(r_B, D_{\rm dest})$\cr 
0 &, $d > \min(r_B, D_{\rm dest})$\cr}~~~.  
\ee
The grain velocity used in evaluating $r_B$ and $D_{\rm dest}$ should be
taken with respect to the ambient ISM.  However, for simplicity, we will
use the velocity with respect to the Sun.

For Local Bubble conditions, magnetic deflection dominates destruction
for grains with $10 < a < 100 \micron$ when $v < 100 \kms$.  
For example, for $a=10 \micron$, $D_{\rm dest} = r_B$ when 
$v=311 \kms$ and $r_B = 281 \pc$ (which exceeds the maximum extent of 
the LB).  For $a=100 \micron$, $D_{\rm dest} = r_B$ when 
$v=143 \kms$ and $r_B = 13 \, {\rm kpc}$.

\section{\label{sec:Vega-like} DUST FROM YOUNG MAIN SEQUENCE STARS}

IRAS observations of several young main 
sequence stars revealed emission at 60 and $100 \micron$ substantially in 
excess of the emission from the star's photosphere (e.g., Aumann et 
al.~1984, Gillett 1986).  This ``Vega phenomenon'' (named for the first
example to be observed) was 
attributed to circumstellar dust, which absorbs the star's optical radiation 
and re-emits in the infrared.  Shortly following this discovery, Smith \& 
Terrile (1984) observed the optical light scattered by the dust around 
$\beta$ Pictoris and found that the grains lie in an edge-on disk.  More 
recently, disks have been imaged (in the infrared and sub-mm)
around a handful of other stars (e.g., Holland et al.~1998, Schneider 
et al.~1999).  For reviews of the Vega phenomenon and circumstellar 
dust disks, see Backman \& Paresce (1993), Lagrange, Backman, \& 
Artymowicz (2000),  and Zuckerman (2001).
It is not yet clear what fraction of Vega-like
stars posses dust disks, but if these stars also posses planets, then 
dynamical interactions can eject the grains from the system.  
(Radiation pressure also removes small grains, but these are too small to
be traced back to their source.)

The most massive disks, which orbit the youngest stars, have optical
depths in the near IR of order $10^{-2}$ to $10^{-3}$. The bulk of
this optical depth is contributed by grains with size
$a_0\sim\lambda/2\pi$, where $\lambda$ is the wavelength, typically
$1-10\micron$ (see below).

\subsection{Gravitational ejection of small particles}
Jupiter mass planets are seen around $5-10\%$ of nearby solar type
stars in radial velocity surveys. These surveys are not sensitive to
planets in orbits larger than $\sim5$ AU, such as Jupiter, so the
fraction of solar type stars with such planets is likely to be
substantially higher. Observations of debris disks also hint at the
presence of planets (Scholl, Roques \& Sicardy 1993; Wilner et
al. 2002). This suggests that gravitational interactions between a
massive planet and dust particles in  debris disks are a natural means for
producing interstellar meteoroids.

We estimate the ejection velocity of large ($25\micron$) dust grains
interacting with a Jupiter mass planet. We neglect collisions with
other dust grains, an assumption which we justify at the end of the
calculation, as well as radiation pressure. We follow the derivation
of \"Opik (1976); a good general introduction to the two body problem
can be found in Murray and Dermott (1999).

We assume that the planet, of mass $M_p$, orbits the star, of mass
$M_*$, on a circular orbit of semimajor axis $\ab_p$. A test mass (the
dust particle) also orbits the star. The test mass is subject to the
gravity of both the planet and the star, but the planet is assumed to
be immune to the gravity of the test particle. When the test particle
is far from the planet it follows a roughly Keplerian orbit around the
star. When the test particle is very close to the planet, it also
follows a roughly Keplerian orbit, but this time around the
planet. While it is close to the planet, the test particle experiences
a tidal force from the star, with a magnitude given by
\be 
F_*\approx {GM_*m_g\over \ab_p^2} {r\over \ab_p},
\ee 
where $r$ is the distance between the test particle and the planet,
$G$ is the gravitational constant, $M_*$ is the mass of the central
star, $\ab_p$ is the semimajor axis of the planet's orbit, and $m_g$ is
the mass of the dust grain. The Hill radius is the distance $r$ at
which this tidal force equals the gravitational force of the planet
on the test particle, $GM_pm_g/r^2$, or
\be \label{eqn: Hill radius}
r_H\equiv\left(M_p\over 3M_*\right)^{1/3}\ab_p,
\ee 
where the factor of $3$ is included for historical reasons.

We make use of the \"Opik approximation; while the test body is inside
the Hill sphere of the planet, we assume that the motion is described
by the two body problem in the frame revolving with the planet. This
involves ignoring both the tidal force of the star and the Coriolis
force associated with the motion of the planet around the star. The
acceleration due to the Coriolis force is
\be 
a_{Cor}\approx \Omega_p V,
\ee 
where $\Omega_p$ is the mean motion of the planet (the angular speed
of the planet's revolution about the star) and $V$ is the
velocity of the test particle in the rotating frame. A rough estimate
for $V$ is $v_p$, the Keplerian velocity of the planet, so the
Coriolis force is larger than the tidal force by a factor $\ab_p/r_H\sim
\mu_p^{-1/3}$ at the Hill radius; in this section $\mu_p\equiv M_p/M_*$.

\subsubsection{Inside the Hill sphere}
We assume that the test particle orbits the star rather than the
planet. Hence the particle follows a hyperbolic orbit relative to the
planet during the close encounter. The hyperbola is specified by its
eccentricity $\bar e$ and the transverse semimajor axis $\bar {\ab}$
(note that the semimajor axis of the dust particle $\bar \ab$ should not be
confused with the radius $a$ of the dust particle). Barred elements
are calculated relative to the planet. Note that $\bar e>1$ and $\bar
\ab<0$. Alternately we can specify the specific energy $\bar
E=-GM_p/2\bar \ab$ and specific angular momentum $\bar L=\sqrt{GM_p\bar
\ab(1-\bar e^2)}$ relative to the planet. The energy is positive, so the
particle would have a finite velocity as it traveled to infinity, if
we ignore the effects of the star. This velocity is known as the
velocity at infinity; we denote it by $U$.

The approximations described above allow us to find an analytic relation
between the angle of deflection resulting from the encounter and the
eccentricity of the (hyperbolic) orbit of the test body around the
planet.  We then relate the eccentricity to the periapse distance
$\bar q$ between the planet and the test particle, and
the velocity at infinity.

The specific energy and angular momentum of a test mass interacting
with a planet of mass $M_p$ are given by 
\be \label{eqn: two body E} 
\bar E={1\over 2 }V^2-{GM_p \over \bar r}
\ee 
and
\be \label{eqn: two body L} 
\bar L=\bar r^2{d\theta\over dt}.
\ee 
In these expressions $G$ is the gravitational constant, ${\bf V}$ is
the velocity of the test mass relative to the planet, and $\bar r$ is
the distance between the two bodies. The angle $\theta$ is measured
from a fixed line (which we choose to be the apsidal line, which
connects the star and the planet at the point of closest approach) and
the line joining the two bodies.

The distance $\bar r$ is given by 
\be \label{eqn: radius} 
\bar r={\bar p \over {1+\bar e\cos\theta}}.
\ee 

The quantity $\bar p\equiv \bar L^2/GM_p$ is called the semilatus
rectum. The periapse distance (at closest approach) is denoted by $\bar
q$ and is given by $\bar q=\bar p/(1+\bar e)$.

The semilatus rectum $\bar p$ is related to $\bar e$ and $\bar \ab$ by
\be \label{eqn: semilatus}
\bar p=\bar \ab(1-\bar e^2)
\ee 
Note that $\bar p>0$, and that $\bar q=\ab(1-e)>0$. Using the definition
of $\bar \ab$, we can write the specific energy in terms of $\bar p$ and
$\bar e$:
\be 
\bar E={GM_p\over 2\bar p}(\bar e^2-1).
\ee 
From eqn. (\ref{eqn: two body E})
\be 
\bar E={1\over2}\bar v_q^2-{GM_p\over \bar q}={1\over2}U^2,
\ee 
where $\bar v_q$ is the speed at closest approach, and $U$ is the
speed at infinity.

From equation (\ref{eqn: radius}), as $r\to\infty$, the angle $\theta$ tends to 
\be 
\theta_0=-\arccos{1\over \bar e}
\ee 
Referring to Figure (\ref{fig:scatter}), and denoting the angle of
deflection by $\gamma$, simple geometry gives
\be 
\sin{\gamma\over 2}=\sin(\theta_0-{\pi\over2})=-\cos\theta_0={1\over
  \bar e}.
\ee 
Since $\bar e=1-(\bar q/\bar \ab)$, we find
\be \label{eqn: gamma}
\sin{\gamma\over 2}=\left[1+{U^2\bar q\over GM_p}\right]^{-1}
\ee 

It is convenient to write this in terms of the impact parameter
$s$. From conservation of angular momentum, $sU=\bar q \bar
v_q$. We find
\be 
\bar q=\sqrt{\left(GM_p\over U^2\right)^2+s^2}-{GM_p\over U^2}
\ee 
Using this in eqn. (\ref{eqn: gamma}) we find
\be \label{eqn: scattering angle}
\sin{\gamma\over 2}=\left[1+{U^4 s^2\over G^2M_p^2}\right]^{-1/2}
\ee 

\subsubsection{Relating $U$ to the orbital elements $\ab$, and $e$ of the
  test particle}

We will assume that the impact parameter $s$ is uniformly distributed
between $0$ and $r_H$. However, we still need to know $U$, the
speed at infinity, in the frame rotating with the planet. We can
find $U$ in terms of the semimajor axis $\ab$ and eccentricity $e$ of
the test particle, where the orbital elements are now calculated
relative to the star. The relations between $E$, $L$, $\ab$, $e$, and
$p$ are as given above, but with the star playing the role of the
central mass, so that  $M_p$ is replaced by $M_*$, the mass of the
star. Since the test particle is bound to the star $E<0$, $\ab>0$, and
$e<1$, at least until the scattering event that ejects the test
particle. We will restrict our attention to the case where the planet
and the test particle orbit in the same plane.

We start by dividing the particle velocity (relative to an inertial
frame centered on the center of mass) into a radial $v_r$ and a
transverse $v_t$ part. From the definition of angular momentum, the
transverse velocity at the time of the close encounter is
\be \label{eqn: v t}
v_t=L/r\approx v_p\sqrt{{\ab\over \ab_p}(1-e^2)},
\ee 
where $v_p=\sqrt{GM_*/\ab_p}$ is the Keplerian velocity of the planet,
and we have made use of the fact that the star-particle distance
$r\approx \ab_p$ (it can differ by $r_H$) during the encounter.

The magnitude of the total velocity $v$ is found from the expression for
the energy,
\be \label{eqn: E}
E={1\over2}v^2-{GM_*\over \ab_p}=-{GM_*\over 2\ab},
\ee 
where we have again set $r=\ab_p$ in the first equality. Solving
equations (\ref{eqn: v t}) and (\ref{eqn: E}) for
$v_r$,
\be 
v_r\approx v_p\left[2-{\ab_p\over \ab}-{\ab\over \ab_p}(1-e^2)\right]^{1/2}.
\ee 
Note that if the orbits just cross, $\ab(1-e)/\ab_p=1$, $1+e=2-\ab_p/\ab$, and
$v_r=0$; the collision occurs at periapse. Then
$U^2=(v_t-v_p)^2\approx v_p^2$. However, the orbit of the test
particle need only pass through the Hill sphere, so that
$\ab(1-e)/\ab_p\approx 1\pm r_H/\ab_p\approx 1\pm \mu_p^{1/3}$, and the radial
velocity $v_r\approx\sqrt{2}\mu_p^{1/6}v_p$, which for a Jupiter mass
planet is about $0.45 v_p$. If there are multiple planets in the
system, the test particle periapse need not be comparable to $\ab_p$, in
which case $v_r$ could be slightly larger than $v_p$. The transverse
velocity $v_t$ is always of order $v_p$.

To find $U$, we transform to the planet frame, which entails
subtracting $v_p$ from $v_t$:
\begin{eqnarray} 
U_r =  v_r     & \approx &\sqrt{2}\mu_p^{1/6}v_p\\
U_t =  v_t-v_p & \approx &v_p\left[\sqrt{{\ab\over \ab_p}(1-e^2)}-1\right]\\
U^2            & \approx & v_p^2\left[2\mu_p^{1/3} + (3-2\sqrt{2})\right].
\label{eq:U}
\end{eqnarray}

\subsubsection{The change in energy}
In the frame rotating with the planet, the close encounter simply
rotates the test particle velocity by the angle $\gamma$, without
changing its magnitude $U$. The components of the scattered velocity
are 
\begin{eqnarray}
U_r' & = & U_r\cos\gamma - U_t\sin\gamma\\
U_t' & = & U_r\sin\gamma + U_t\cos\gamma.
\end{eqnarray}

Transforming back to the inertial frame, the square of the new
velocity is given by
\be 
v'^2=U_r'^2+(U_t'+v_p)^2=U_r^2+U_t^2+2v_p(U_r\sin\gamma+U_t\cos\gamma)+v_p^2.
\ee 
The change in specific energy  
\be 
\Delta E \equiv {(v'^2-v^2)\over2} \approx v_p
\left[U_r\sin\gamma + U_t(1-\cos\gamma)\right]
\ee 
The radius $r$ does not change
appreciably during the encounter.
Using the relations between $U$ and $v$,
\be 
\Delta E = v_p\left[v_r\sin\gamma
+(v_t-v_p)(1-\cos\gamma)\right].
\ee 

\subsubsection{The small scattering angle approximation}
Equation (\ref{eq:U}) shows that $U^2\approx 0.4v_p^2$ for test
particles with $\ab>>\ab_p$ scattering off Jupiter mass planets.  From
equation (\ref{eqn: scattering angle}), we find
\be \label{eqn: small angle}
\gamma\approx 2GM_p/(sU^2) = 2{\root 3\of 3}\mu_p^{2/3}
\left({v_p\over U}\right)^2\left(r_H\over s\right).
\ee 
%
%
%
The small angle approximation is valid for $s\gtsim s_\gamma$, where
\be 
s_\gamma\approx\mu_p \left(v_p\over U\right)^2\ab_p
\ee 

There is another constraint on the minimum value of the impact
parameter $s$, namely that the test particle does not physically
collide with the planet. This defines $s_{min}$, the impact parameter
for which the periapse $\bar q=r_p$, where $r_p$ is the radius of the
planet. For $\ab>>\ab_p$, $s_{min}\approx r_p\sqrt{2\mu_p(v_p/U)^2 \ab_p/r_p}$. At 5AU
$s_{min}\approx 7 r_p$ for a planet with Jupiter's mass and
radius. Thus non-collisional close encounters occur if the impact
parameter satisfies
\be 
7r_p\lesssim s\lesssim r_H,
\ee 
For a Jupiter mass planet at 5 AU, $s_\gamma\approx 25 r_p$,
about four times larger than the minimum impact parameter. For the
same values of $M_p$ and $\ab_p$, $r_H\approx 750r_p$. In other words,
the small angle approximation is valid for almost any impact parameter
that does not lead to a physical collision.

Assuming $\gamma<<1$, the change in energy due to a single close
encounter for a test particle with $\ab>>\ab_p$ is
\be \label{eqn: delta E}
\Delta E(\mu_p,\ab_p,s)\approx 2^{3/2}{\root 3\of 3}\,\mu_p^{5/6}
\left(v_p\over U\right)^2
\left(r_H\over s\right)v_p^2.
\ee 

\subsubsection{The ejection velocity}

We are now in position to estimate the typical ejection velocity of a
test particle interacting with a Jupiter mass planet. The particle
must have an energy satisfying $-\Delta E<E<0$ when the last encounter
occurs. On average it will emerge with an energy $E\approx \Delta E/2$
after the encounter, so it is ejected with a velocity
\be \label{eqn: kick}
v_\infty(\mu_p,\ab_p,s)\approx2\mu_p^{5/12}\left(v_p\over U\right)\sqrt{r_H\over s}v_p.
\ee 
Note that the ejection velocity scales as the $1/2$ power of the
planet's semimajor axis $\ab_p$. The scaling with the mass ratio $\mu_p$
is complicated by the appearance of the Safronov number $U/v_p$. From
equation (\ref{eq:U}), the Safronov number is independent of $\mu_p$ for
$\mu_p<<6\times10^{-4}$, but scales as $\mu_p^{1/6}$ for larger
$\mu_p$. Hence the ejection velocity will scale as $\mu_p^{5/12}$ for
Saturn mass planets, and as $\mu_p^{1/4}$ for planets  significantly
more massive than Jupiter. If the planet has a substantial
eccentricity, as many extrasolar planets do, then $v_t\approx v_p$ and
the  ejection velocity will scale as $\mu_p^{1/3}$.

For a Jupiter mass planet at $5$ AU from a solar mass star,
\be \label{eqn: Jupiter kick}
v_\infty(\mu_p,\ab_p,s)\approx 2\left(\mu_p \over10^{-3}\right)^{1/3}
\left(5{\,\rm AU}\over \ab_p\right)^{1/2}
\left(v_p\over U\right)
\sqrt{r_H\over s}{\km/\s}.
\ee 
Let $x=s/r_H$; then the cross section for an encounter with impact
parameter between $s$ and $s+ds$ is $2\pi sds=2\pi r_H^2\, xdx$.  We
average over $x$ to find the mean escape velocity. We integrate from
$x_{\gamma}\equiv s_{\gamma}/r_H \approx \mu_p^{2/3}<<1$ to $s=\alpha
r_H$, where $\alpha$ is a dimensionless constant of order unity, and
$s_{\gamma}$ is the minimum value of $s$ for which equation (\ref{eqn:
small angle}) is valid. Doing the integration,
\be 
\left<v_\infty\right>={\int_{x_\gamma}^\alpha \pi x v_\infty(x) dx\over 
\int_{x_\gamma}^\alpha \pi x dx}\approx {4v_\infty(r_H)\over3\sqrt{\alpha}},
\ee 
where we have neglected terms of order $\mu_p v_\infty(r_H)$; we use the
notation $v_\infty(r_H)=v_\infty(\mu_p, \ab_p, s=r_H)$. 
The rms escape velocity is calculated in a similar manner. We find
\be \label{eqn: v_rms}
v_{rms}\approx \sqrt{2\over \alpha}v_\infty(r_H),
\ee 
so the spread in escape velocities is similar to the mean escape velocity.

For those rare
encounters with $s_{min}<s<s_\gamma$, the scattering angle is of order
unity, and the velocity kick experienced by the particle is larger
than the estimate in eqn. (\ref{eqn: kick}). We take
$\sin\gamma\approx\cos\gamma\approx 1/\sqrt{2}$, so 
\be 
\Delta E\approx v_p^2,
\ee 
and
\be 
v_\infty\approx v_p,
\ee 
much larger than the small angle limit. About 1\% of the particles
will be ejected with velocities of tens of kilometers per second.

\subsubsection{Ejection time scale compared to collisional timescale}
So far we have assumed that the grains are ejected before they suffer
sufficient collisions with other dust particles to alter their orbits
substantially. We are now in a position to check this assumption. We
start with an estimate of the dust-dust collision time. The optical
depth $\tau$ for the most massive debris disks is of order
$0.001$. This optical depth is due to particles with radii
$a\approx\lambda/2\pi$, where $\lambda$ is the wavelength at which the
disk is observed. Since the disks are usually detected by their IR
excesses, the dust particles responsible for the optical depth have
$a\approx0.3\micron$ or smaller. We are interested in larger ejected
particles, with $a_{eject}=25\micron$. These larger particles will
suffer a collisions with a particle of size $a_{target}$ after roughly
$1/\tau(a_{target})$ orbits, where $\tau(a_{target})$ is the optical
depth to particles of size $a_{target}$.

Using the scaling $mf_m\propto m^{-1}$, the optical depth in particles
of size $a_{target}$, as viewed at a wavelength of order or smaller
than $a_{target}$, is
\be 
\tau(a_{target})=\left(0.3\micron\over a_{target}\right)\tau(a=0.3\micron).
\ee 

The typical number of orbits required for our test (ejected) particles
to sweep up their own mass in smaller particles, and hence have their
orbits altered substantially, or to strike a larger particle, is
\be 
N \approx 
{1\over\tau(a=0.3\micron)}
\Bigg\{
\begin{array}{cc}
\left({a_{eject}\over 0.3\micron}\right)
\left(a_{eject}\over a_{target}\right)^2, & a_{target}\le a_{eject}\\
\left({a_{target}\over 0.3\micron}\right), & a_{target}\ge a_{eject}
\end{array}
\phantom{\}}
\ee 
orbital periods. 

Thus the most efficient way to alter the orbit of a particle is to
collide with another particle of the same size, assuming that the mass
is logarithmically distributed, and assuming that gravity plays no
role in the collision. The typical number of orbits for a $25\micron$
size particle to collide with a particle of its own mass is
$\sim1/\tau(a=25\micron)\approx7\times10^4$.

Dust-dust collisions can also shatter the grains. The binding energy
for a target grain of mass $m$ is $\zeta m$. A collision with a grain of mass
$m_c$ moving with relative velocity $v$ will shatter the target if the
kinetic energy exceeds the binding energy of the target. Allowing for
the possibility that only a fraction $f_{KE}$ of the kinetic energy is
available to disrupt the target, the critical mass needed to shatter
the target is
\be 
m_c={2\over f_{KE}}{\zeta\over e^2v_p^2}m.
\ee 
In deriving this result we assume that the relative velocity is
$ev_p$, corresponding to the random velocity of a grain suffering
close encounters with a massive planet of semimajor axis $\ab_p$. The
eccentricity of the target grain undergoes a random walk in $e$,
starting at zero and evolving to $e=1$, $e(t)\approx \sqrt{D_e t}$, so
we take $e=2/3$. The minimum mass needed to shatter the target grain
is then $m_c\approx 0.5(\ab_p/5 {\rm AU})m/f_{KE}$, roughly a mass equal
to that of the target. 

How long does it take for the typical particle to be
ejected? The test particles undergo a random walk in energy, with a
step size given by equation (\ref{eqn: delta E}). The number of
collisions needed to random walk from $E=-GM/2\ab_p=-v_p^2/2$ to $E=0$ is
$N\approx 3^{2/3}/16\mu_p^{4/3}$. The probability of a close encounter on
each periapse passage is 
\be 
P\approx {\pi r_H^2\over(2\pi \ab_p r_H)}\approx{1\over 2}
\left({\mu_p\over 3}\right)^{1/3}.
\ee 
The total number of orbits up to ejection is 
\be 
N_{eject}\approx{3\over8}\mu_p^{-5/3}\approx4\times10^4
\left(10^{-3}\over\mu_p\right)^{5/3}.
\ee 
We conclude that particles larger than about $25\micron$ will be
ejected before their orbits are significantly altered by collisions
with smaller or similar size dust grains.

\subsubsection{Numerical Results} 
We have carried out numerical integrations of test particles in the
gravitational field of one or more massive planets orbiting a star. We
use the publicly available SWIFT (Levison \& Duncan 1994)
integration package, which is based on the Wisdom \& Holman (1991)
symplectic integration scheme. We used the rmvs3 integrator, in order to
integrate through close encounters between the test particles and the
planets. All integrations were for 100 million years, with a timestep
chosen to be small enough to resolve periapse passage for the
estimated most extreme test particle orbits, or to resolve the
perijove passage for planet grazing close encounters (taken to be at
$2r_p$), whichever was smaller. If a particle passed within $2r_p$ is
was deemed to have collided with the planet and was removed from the
integration.

The initial eccentricities and inclinations (measured from the
planet's orbital plane) of the test particles were generally set to
$0.1$ and $0.087$ radians respectively, although we tried runs with
other values. The final ejection velocities did not depend strongly on
the initial $e$ and $i$. The test particles were given semimajor axes
ranging from $0.5$ to $1.5\ab_p$ in single planet cases, since in those
cases particles starting at larger distances from the planet were
typically not ejected by the time the integrations were halted.

Figure (\ref{fig:swift}) shows the result of a numerical integration
of $\sim 600$ test particles in the gravitational field of Jupiter and
the Sun. Our analytic calculations should be a good approximation to
this case, since $e_J=0.048$. The test particles were started with a
range of semimajor axes between $x$ and $y$, with initial
eccentricities $e\sim0.05$ and inclinations $i\sim0.05$ radians. The
figure shows the velocities at which the particles were ejected,
including the (small) correction for their finite distance from the
Sun when the integration was stopped. The stopping criterion was that
the test particle have a positive energy relative to the Sun and a
semimajor axis larger than $100$AU. The mean escape velocity is
$\sim1\km\s^{-1}$, slightly lower than predicted by equation
(\ref{eqn: Jupiter kick}). 

We found that the final inclinations were small, typically within $10$
degrees of the orbital plane of the planet, but with a significant
minority of particles ejected at inclinations up to $30$ degrees.

We checked the scaling of $v_{ej}$ with $\ab_p$ by varying the size of
the planet's orbit over a decade. A least squares fit to the mean
ejection velocity as a function of $\ab_p$ gives $v_{ej}\sim \ab_p^{-0.5}$
with an error of about $0.05$ in the exponent. This is consistent with
the predicted slope of $-0.5$ given by equation (\ref{eqn: kick}). We
also checked the scaling with $M_p$, finding $v_{ej}\sim M_p^{0.28}$,
compared with the prediction exponent of $1/4$. In summary, equations
(\ref{eqn: kick}) and (\ref{eqn: Jupiter kick}) appear to be a good
description of the ejection process.

The numerical integrations employing multiple Jupiter mass planets
gave similar results, with the test particle ejection velocity
determined by the most massive of the planets in the integration. The
same was true of integrations involving a single Jupiter mass planet
together with several 1-10 Earth mass bodies.

These results suggest that equations (\ref{eqn: Jupiter kick}) and
(\ref{eqn: v_rms}) can be used to describe the ejection velocity of
small particles in systems with one or more Jupiter mass planets.

\subsection{Dust Luminosity}

In order to estimate the (age-dependent)
dust luminosity of a Vega-like star, we must 
first estimate the number of grains in the disk.  Spangler et al.~(2001)
have measured IR excesses for stars in several nearby clusters.  Their 
results are consistent with the following relation:  dust disk mass 
$\propto$ (age)$^{-2}$.  We normalize this relation by estimating the 
(size-dependent) number of grains in the disk around $\beta$ 
Pictoris, with an age $\approx 12 \Myr$ (Zuckerman et al.~2001).

We assume that the grains around $\beta$ Pic are located at a distance 
$D = 100 {\rm AU}$ from the star  
and that they have a size distribution $dN/da = C a^{-3.5}$, where
$N(a)$ is the number of grains with size $\le a$.  This is the equilibrium
size distribution that results when mass is redistributed among grain sizes
via shattering in collisions (Dohnanyi 1969).  We obtain the normalization
constant $C$ by setting the emission at $800 \micron$ equal to the observed
value of $F_{\nu} = (115 \pm 30) {\rm mJy}$ (Zuckerman \& Becklin 1993).
The specific flux is given by
\be
\label{eq:F_nu}
F_{\nu} = \frac{\pi}{D^2} \int da \frac{dN}{da} \Qabs a^2 B_{\nu}[T(a)]~~~,
\ee
where the absorption cross section is $\Qabs \pi a^2$, $B_{\nu}(T)$ is the
Planck function, and the grain temperature $T(a)$ is determined by 
equating the absorption and emission rates.  For $\beta$ Pic, we take
effective temperature $T_{\rm eff} = 8200 \K$, luminosity 
$L = 8.7 L_{\odot}$, and radius $R = 1.47 R_{\odot}$ (Crifo et al.~1997).
Grains with $a \gtsim 1 \cm$ contribute little to $F_{\nu}(800 \micron)$.
Terminating the integral in equation (\ref{eq:F_nu}) at $a=1 \cm$ 
yields $C = 4.65 \times 10^{25} \cm^{2.5}$ for silicate grains, implying 
a disk mass of $1.4 \times 10^{27} \g$ ($\approx 19$ lunar masses, 0.23
Earth masses) in
grains with $a \le 1 \cm$.  If we assume graphite (water ice) rather than 
silicate composition, then the estimated mass increases by a factor of
1.6 (4.6).  Since pure water ice grains are very unlikely, our mass 
estimate is robust against variations in compositions.  If the grains are
located $30 \pc$ ($500 \pc$) rather than $100 \pc$ from the star, then the 
estimated mass decreases by a factor of 2 (increases by a factor of 3).
Thus, though the dust actually will be distributed over a range of distances
from the star, this does not seriously affect the estimate of the total
dust mass.

Our result can be compared with the analysis of Li \& Greenberg (1998),
who constructed a detailed dust model to account for the emission (at 
multiple wavelengths) from the dust around $\beta$ Pic, using fluffy grains.  
In their model, the total dust mass needed to produce the observed 
emission is $\sim 2 \times 10^{27} \g$, lying mostly beyond 100 AU from 
the star.

It is interesting to note that estimates of the dust mass in circumstellar
disks typically assume that the grains are in the Rayleigh limit, i.e., 
that $a \ll \lambda$, the wavelength of the observation (see, e.g., the 
review by Zuckerman 2001).  In this regime, $\Qabs \propto a$.  In the 
opposite regime ($a \gg \lambda$), $\Qabs \sim 1$.  Suppose $a_0$ is 
the minimum $a$ for which $\Qabs \sim 1$, for fixed $\lambda$.  Then,
$\Qabs \sim a/a_0$ in the Rayleigh limit.  The opacity $\kappa$ 
(absorption cross section per unit mass) of the emitting grains is 
the crucial quantity for estimating dust mass from emission; 
$\kappa = 3 \Qabs / 4 a \rho$.  In the Rayleigh limit, 
$\kappa \sim 3 / 4 a_0 \rho$.  In our estimate of the dust mass (with our
assumption that $dN/da \propto a^{-3.5}$), the emission is dominated by 
grains with $a \sim a_0$, since these are the most abundant grains with 
$\Qabs \sim 1$.  Thus, $\kappa \sim 3 / 4 a_0 \rho$ for our estimate as
well as for the Rayleigh limit, and these two different methods yield 
approximately the same mass in grains contributing significantly 
to the observed 
emission.  We feel that our scenario, with a Dohnanyi size distribution,
is more realistic than the standard scenario, in which all of the grains
that contribute significantly to the emission are supposed to be in the 
Rayleigh limit.  

Combining the above estimate for the number of grains in the disk around
$\beta$ Pic with the Spangler et al.~(2001) result, we estimate 
\be 
\frac{dN}{da} \sim 4.7 \times 10^{25} \cm^{2.5} \left( \frac{12 \Myr}{t}
\right)^2 a^{-3.5}~~~,~~~~~a \ltsim 1 \cm~~~,
\ee 
where $t$ is the age of the system.  

The Spangler et al.~(2001) result suggests that grains are lost at a rate
$dN/dt \sim -2 N/t$.  The following could be important sinks for the 
grains:  1.  gravitational ejection from the system,  2.  gravitational 
ejection into the star,  3.  incorporation into planets or 
planetesimals,  and  4.  ejection of small shattering fragments by 
radiation pressure.  We will not attempt to estimate the relative importance
of these sinks in this paper.  Rather, we will introduce the unknown 
factor $f_{\rm ej}$, equal to the fraction of the grains that are lost by 
gravitational ejection.  Then, the specific dust luminosity is 
\be
L_{v, a}(t, \vej, a) = 7.8 \times 10^{18} \yr^{-1} \cm^{-1} (\cm/\s)^{-1} 
\left( \frac{12 \Myr}{t}\right)^3 \left( \frac{a}{\cm} \right)^{-3.5} 
f_{\rm ej} f_v(\vej)~~~,
\ee
where we have taken time equal to the age of the star when the grains are
ejected and 
$f_v(\vej) d\vej$ is the fraction of the grains ejected with speed
between $\vej$ and $\vej + d\vej$.\footnote{We assume that $f_v$ is 
independent of $a$.}  

At very early ages, the primordial gas disk will not yet have dissipated,
and the drag force on grains could dramatically reduce the dust 
luminosity.  Also, we do not expect large grains to be ejected prior to 
planet formation.  Thus, $L_{v, a} = 0$ for $t < t_{\rm cr}$, an unknown 
critical age.  We will assume that $t_{\rm cr}$ lies in the range 3 to 
$10 \Myr$.

\subsection{Flux at Earth}

Assuming $f_{\rm survive} = 1$, 
the flux, at Earth, of grains with size $\ge a$, $F(a)$, is given by
\be
\label{eq:flux_at_earth}
F(a) = 8.2 \times 10^{-5} \yr^{-1} \, {\rm km}^{-2} f_{\rm beam}
f_{\rm ej} \left( \frac{d}{10\pc}
\right)^{-2} \left( \frac{a}{10\micron} \right)^{-2.5}
\int d\vej \, f_v(\vej) \frac{v_{d, \odot}}{\vej} Q(t_{\rm ej} = t - 
d/\vej)~~~,
\ee
where
\be
Q(t_{\rm ej}) = \cases{(12 \Myr/t_{\rm ej})^3 &, 
$t_{\rm ej} \ge t_{\rm cr}$\cr
0 &, $t_{\rm ej} < t_{\rm cr}$\cr}~~~.
\ee 

If we assume $v_{d,\odot}/v_{ej}\approx10$, this yields a flux of
$2.6\times10^{-21}(25\micron/a)^{2.5}(10\pc/d)^2\cm^{-2}\s^{-1}$,
which may be compared with Fig. (\ref{fig:flux}); the AMOR point
source supplies a flux at Earth of $10^{-17}\cm^{-2}\s^{-1}$. It would
appear that the point source is not due to a debris disk.

We will be interested in sources within the
LB, so $f_{\rm survive}=1$ when $d < r_B$.  With $\Phi \approx 0.94 \,
{\rm V}$ and $\rho = 3.5 \g \cm^{-3}$ (appropriate for silicate grains
in the LB), equation (\ref{eq:gyroradius}) yields a critical grain
radius \be
\label{eq:a_cr}
a_{\rm cr} \equiv 10 \micron \left( \frac{d}{9.0 \pc} \right)^{1/2}
\left( \frac{v}{10 \kms} \right)^{-1/2}~~~;
\ee
$f_{\rm survive} = 1$ when $a > a_{\rm cr}$.  We will use 
$v_{\ast, \sun}$ as a good approximation to $v_{d, \sun}$ in equation 
(\ref{eq:a_cr}).  

Suppose the threshold of detectability requires 20 events per year, so
that the threshold flux $F_{\rm th}=20 \yr^{-1} A_{col}^{-1}$.  Then,
we can estimate the threshold distance $d_{\rm th}$ out to which
sources might be detected by making the optimistic assumptions that
$t_{\rm ej} = t_{\rm cr}$ and $f_{\rm ej}=1$ (we will also assume that
$f_{\rm beam}=1$).  Inserting equation (\ref{eq:a_cr}) into equation
(\ref{eq:flux_at_earth}) and assuming $v_{d, \odot}/\vej \sim 10$
yields 
\be 
d_{\rm th} \approx 0.42 \left( \frac{A_{col}}{1 \km^2}
\right)^{4/13} \left( \frac{t_{\rm cr}}{12 \Myr} \right)^{-12/13}
\pc~~~.  
\ee
In Table \ref{tab:d_th} we give $d_{\rm th}$ and $a_{\rm cr}$ for the
cases that $T_{\rm cr} = 3$ or $10 \Myr$ and $A_{col} = 10^4 \km^2$ or
$10^6\km^2$; recall that AMOR has $A_{col}\approx10\km^2$.

\begin{figure}
\epsscale{1.00}
\plotone{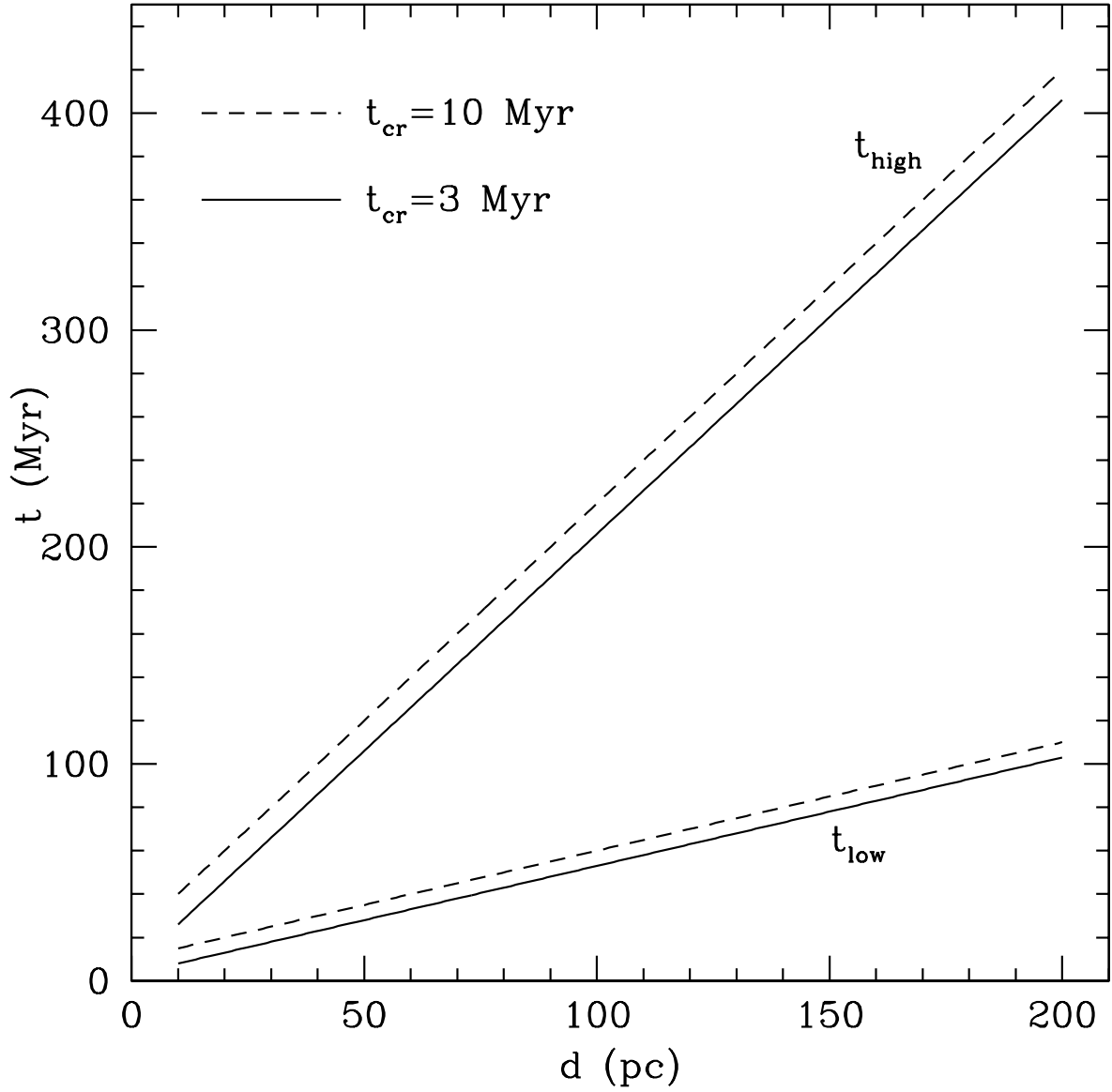}
\caption{
\label{fig:age_ranges}
The upper and lower current ages $t$ of a Vega-like star for which we would 
observe grains emitted when the star's age was between $t_{\rm cr}$ 
and $2 t_{\rm cr}$, versus the distance $d$ to the star.   
        }
\end{figure}

\begin{figure}
\epsscale{1.00}
\plotone{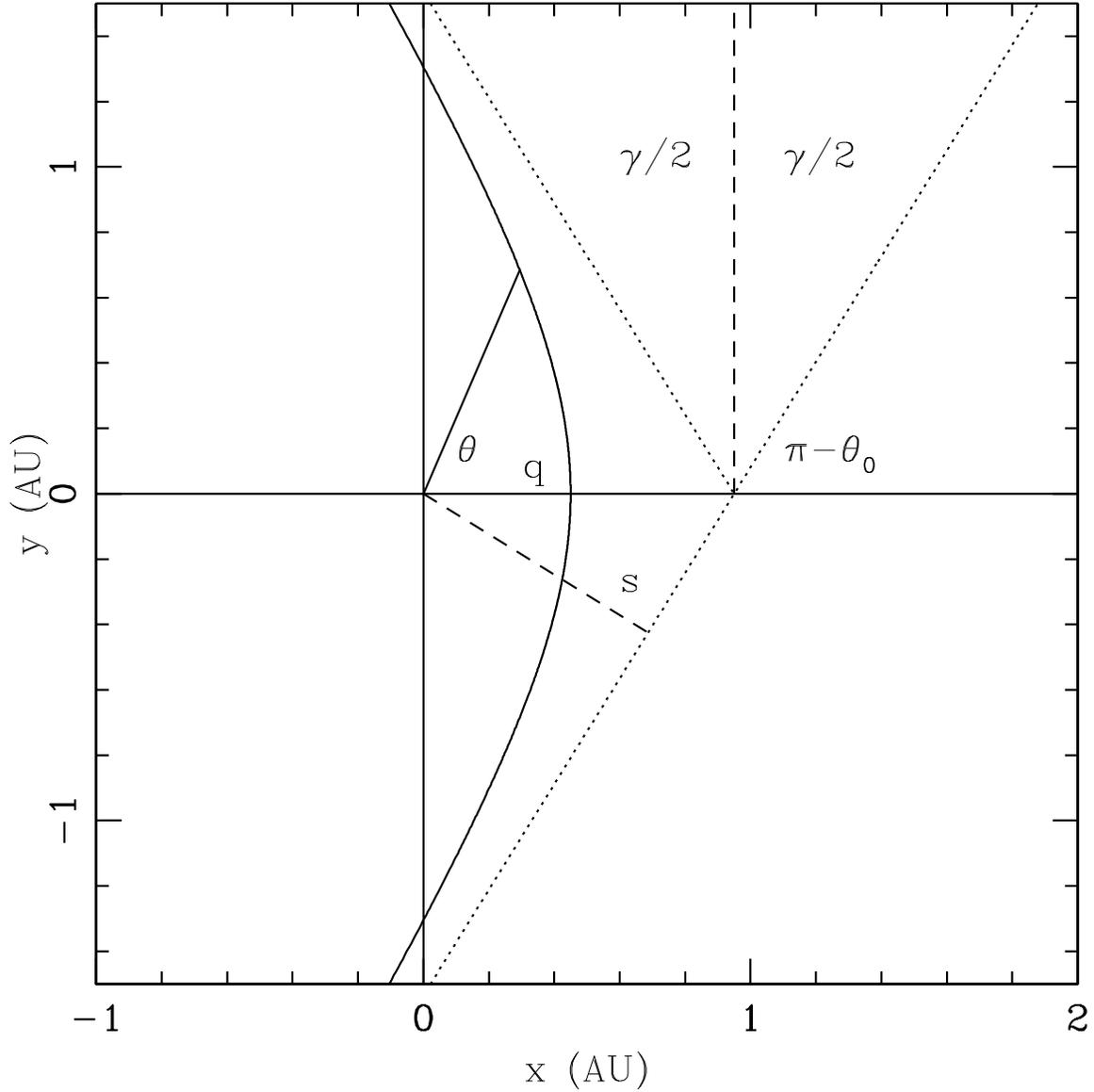}
\caption{ The geometry of hyperbolic motion in the two body
 problem. The scattering center (the planet) is at the origin; the
 test particle travels along the curved line, with an initial impact
 parameter given by $s$. Its closest approach to the planet occurs as
 it crosses the apsidal line (choosen here to be the x-axis), when it
 is a distance $q$ from the planet. The ingoing and outgoing
 asymptotes are denoted by dotted lines. The angle between the ingoing
 and outgoing asymptotes is $\gamma$.
\label{fig:scatter}
}
\end{figure}

\begin{figure}
\epsscale{1.00}
\plotone{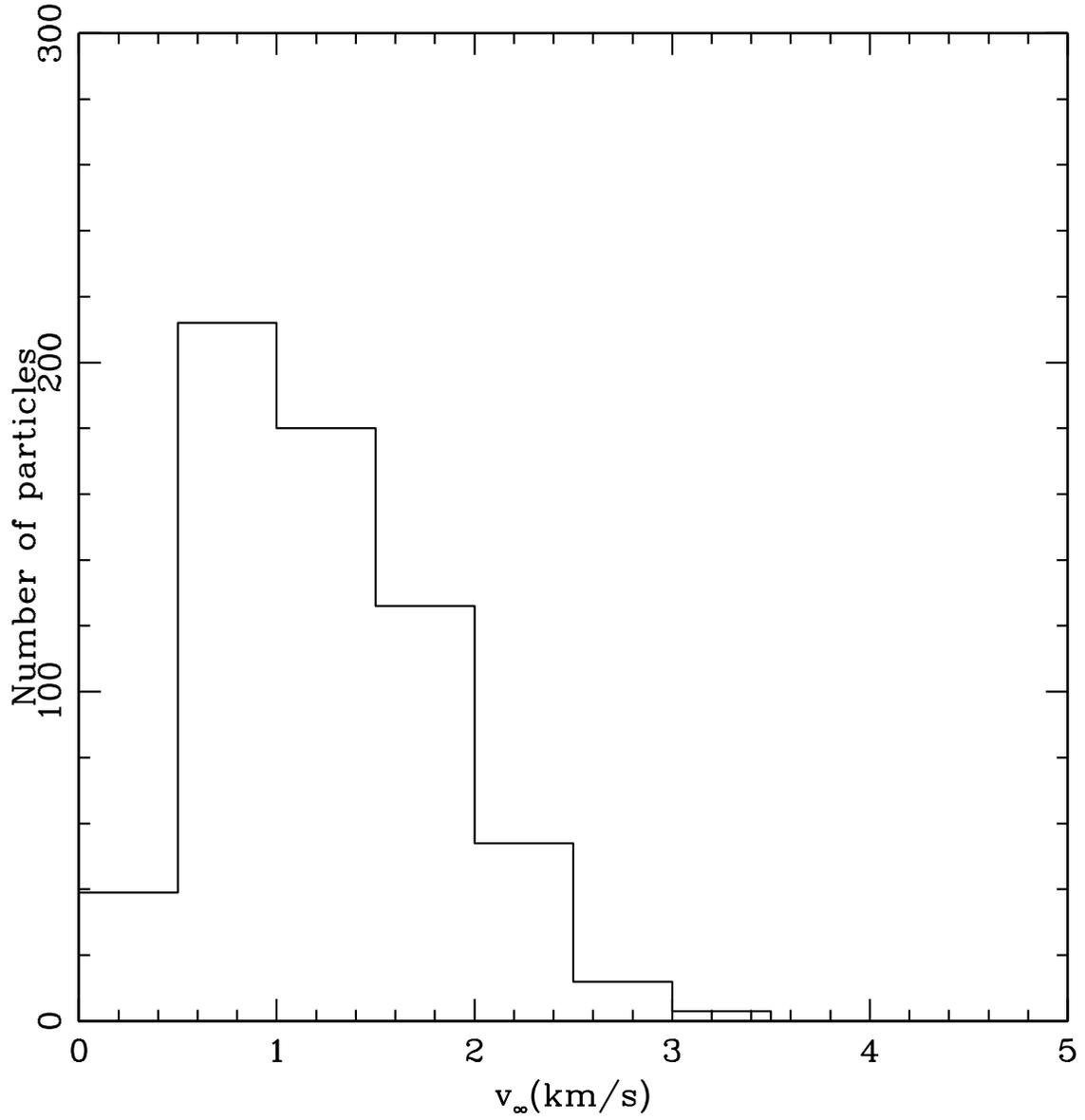}
\caption{
\label{fig:swift}
Ejection velocities of small particles from a system with a solar mass
star and a Jupiter mass planet at $5.2$AU.
}
\end{figure}

Since the flux drops off as $t_{\rm ej}^{-3}$, it is highly advantageous
to catch the star when $t_{\rm ej} \sim t_{\rm cr}$.  In Figure
\ref{fig:age_ranges}, we plot, as a function of $d$, the upper and lower 
ages for which $t_{\rm cr} \le t_{\rm ej} \le 2 t_{\rm cr}$, assuming a
range of ejection speeds between $0.5 \kms$ and $2 \kms$.  Since the 
total volume and favorable age range both increase with $d$, distant
stars that are possibly too old to have observable IR excesses today 
may be a significant source population, despite the $d^{-2}$ decrease 
in flux.  However, nearby identifiable Vega-like stars are more 
attractive sources, since we could then combine the dust flux information
with other observations to learn more about a particular object.

\subsection{Candidate Sources}

\subsubsection{\label{sec:Gliese} Gliese Catalog Stars}

Are there any nearby Vega-like stars for which we can expect a detectable
dust flux at Earth?  To answer this question, we consider stars from the 
Gliese Catalog with far-infrared excesses from IRAS (see Table X in 
Backman \& Paresce 1993).  

In Table \ref{tab:Gliese}, we give each star's distance $d$ and its space
velocity $(U,V,W)$ with respect to the Sun.  In calculating these quantities,
we have taken coordinates, parallaxes, proper motions, and radial velocities
from SIMBAD.\footnote{Note that we exclude one star from Table \ref{tab:Gliese} in 
Backman \& Paresce (1993), Gl 245, because SIMBAD does not give its 
radial velocity.}  We estimate a star's age $T$ from its IR excess $\tau$,
which is the fraction of the star's luminosity that is re-emitted in the 
IR by dust.  Adopting the Spangler et al.~(2001) result that 
$\tau \propto t^{-2}$, we take
\be
\label{eq:T_tau}
t = t(\beta \, {\rm Pic}) \left[ \frac{\tau(\beta \, {\rm Pic})}{\tau}
\right]^{0.5}~~~.
\ee

The dust fluxes in Table \ref{tab:Gliese} were calculated using
equation (\ref{eq:flux_at_earth}) with limiting grain size $a=\max(10
\micron, a_{\rm cr})$ ($a_{\rm cr}$ is taken from eq.~\ref{eq:a_cr}
with $v = v_{\ast, \sun}$), a flat distribution of ejection velocities
between $0.5 \kms$ and $3 \kms$, $t_{\rm cr} = 3 \Myr$, and $f_{\rm
beam} = f_{\rm ej} =1$.  Thus, these estimates are optimistic.  In \S
\ref{sec:Discussion} below, we suggest how to build a radar detector
with $A_{col}\approx3\times10^4\km^2$. Four of the stars in Table
\ref{tab:Gliese} would yield 20 or more meteors per year (i.e., $F > 6
\times 10^{-4} \yr^{-1} \km^{-2}$) for such a system.  Since $t_{\rm
ej}$ is substantially greater than $t_{\rm cr}$ for most of the stars,
increasing $t_{\rm cr}$ to $10 \Myr$ yields reduced fluxes for only
three stars: Gl 219, 297.1, and 673.1.  The flux for Gl 219 ($\beta
\,$Pic) vanishes, whereas the fluxes for Gl 297.1 and 673.1 remain
greater than $2 \times 10^{-5} \yr^{-1} \km^{-2}$.

The apparent location of a dust source on the sky is determined by the velocity
of the dust relative to the Sun (eq.~\ref{eq:v_dust_Sun}).  If the dust
ejection speed $\vej$ were much greater than the star's speed 
$v_{\ast, \sun}$, then the dust would
appear to come from the location of the star itself (since the speed of 
light is much greater than the star's speed).  However, $\vej$ is 
typically an order of magnitude or more smaller than $v_{\ast, \sun}$.
Thus, the location of the dust source on the sky is primarily determined by 
$\vec{v}_{\ast, \sun}$ and need not be anywhere near the actual location of 
the source.  

\begin{figure}
\epsscale{1.00}
\plotone{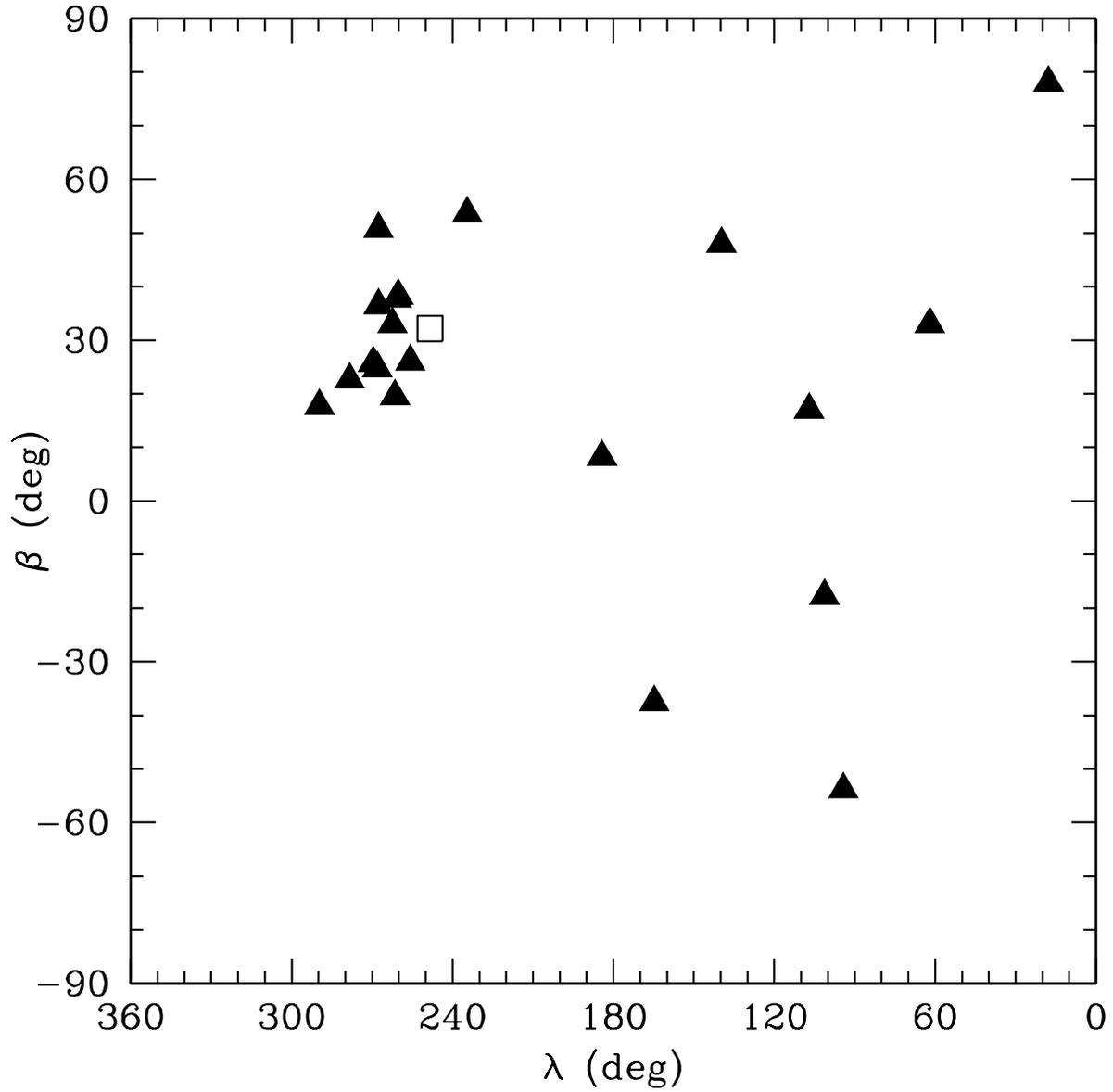}
\caption{
\label{fig:Gliese}
Triangles:  apparent locations of the dust flux from the Gliese stars in 
Table \ref{tab:Gliese}.  Open box:  Direction of the solar apex.  
Coordinates are ecliptic longitude ($\lambda$) and latitude ($\beta$).
        }
\end{figure}

Since $\vec{v}_{\ast, \sun}$ is determined in part by the velocity 
$\vec{v}_{\sun, {\rm LSR}}$ of 
the Sun relative to the local standard of rest (LSR), we may expect some 
degree of clustering of the apparent directions around the solar apex 
(i.e., the direction of $\vec{v}_{\sun, {\rm LSR}}$).  
Dehnen \& Binney (1998) give $U=10.00 \pm 0.36$, $V=5.25 \pm 0.62$, and
$W=7.17 \pm 0.38 \kms$ for $\vec{v}_{\sun, {\rm LSR}}$.  They find that the
velocity of the Sun with respect to nearby young stars is similar, except
that $V \approx 12 \kms$ in this case.  The coordinates of the solar 
apex with respect to the LSR are $(l, b) = (27.7\arcdeg, 32.4\arcdeg)$
or $(\lambda, \beta) = (248.4\arcdeg, 32.2\arcdeg)$, while those with
respect to nearby young stars are $(l, b) = (50.2\arcdeg, 24.7\arcdeg)$
or $(\lambda, \beta) = (265.7\arcdeg, 48.9\arcdeg)$.  In addition to 
Galactic coordinates $(l, b)$, we also make use of ecliptic coordinates:
ecliptic longitude $\lambda$ and ecliptic latitude $\beta$.  These 
coordinates are useful in radar studies of meteors, because most of the 
observed grains originate in the ecliptic plane.  
In Table \ref{tab:Gliese}, we give the apparent direction to the dust 
stream in ecliptic coordinates.  Figure \ref{fig:Gliese} is a plot of 
these directions on the sky.

\subsubsection{Young Clusters}

Nearby young clusters could potentially yield strong dust fluxes.  The 
cluster members may not have appeared in searches for the Vega Phenomenon,
since the stellar luminosities may be too low.  However,
most nearby clusters have poor tuning between distance and age.  There are 
some very young clusters that are too far away for the dust to have reached 
us yet and some nearer clusters that are much older (see Table 1 in 
Spangler et al.~2001).  

The recently discovered Tucana Association 
(Zuckerman, Song, \& Webb 2001) might be more suitable.  Zuckerman \& 
Webb (2000) find a distance $\sim 45 \pc$, age $\sim 40 \Myr$ and space
velocity $(U,V,W) \approx (-11, -21, 0) \kms$.  Stelzer \& Neuh\"{a}user
(2000) find a younger age of 10--30$\, \Myr$.  Adopting numbers from 
Zuckerman \& Webb and making use of the convenient coincidence that 
$1 \kms = 1.02 \pc/\Myr$, we find that dust ejected at speed
$\vej = 1.5 \kms$ when the stars were $15 \Myr$ old would be reaching us 
today.  The position of the dust stream on the sky would be 
$(\lambda, \beta) = (126 \arcdeg, -46 \arcdeg)$.  

The Pleiades cluster is another potential source, with a distance of 
$118 \pc$ and an age of $\approx 120 \Myr$ (Spangler et al.~2001).
Although the large distance would suppress the flux (because of both the 
$d^{-2}$ dependence in eq.~\ref{eq:flux_at_earth} and the need for 
larger grains in order to avoid magnetic deflection), this could be 
partially compensated by the large number of stars in the cluster.
Robichon et al.~(1999) find $(U, V, W) = (-6.35, -24.37, -13.02) \kms$
for the cluster motion.  Ignoring the Galactic potential, this would yield
$(\lambda, \beta) \approx (276\arcdeg, 73\arcdeg)$ for the position of the 
dust stream.

\subsection{Has Dust from $\beta \,$Pic Been Detected?}

Baggaley (2000) detected a ``discrete'' source of radar meteoroids with a 
central location $(\lambda, \beta) \approx (280 \arcdeg, -56 \arcdeg)$, an 
angular diameter of $\sim 30 \arcdeg$, and a dust speed relative to the 
Sun $v_{d, \sun} \approx 13 \kms$.  He claimed that the central location
and dust speed could be reproduced if $\beta \,$Pic were the source and 
the grains were ejected with speed $\vej \approx 29 \kms$.  

This result seems unlikely on theoretical grounds, since it is not
clear how the grains can be ejected with such large speeds.  In
addition, we do not find that dust emitted from $\beta \,$Pic
reproduces the location of the discrete source for any $\vej$.
Baggaley did his calculations in the LSR frame, taking the Sun's
motion with respect to the LSR from Binney \& Tremaine (1987), who
give $(U_{\sun}, V_{\sun}, W_{\sun}) = (9, 12, 7) \kms$.  Baggaley
found the direction to the discrete source to be $(\lambda, \beta) =
(49\arcdeg, -72\arcdeg)$, $(l, b) = (259\arcdeg, -28\arcdeg)$, whereas
we find $(\lambda, \beta) = (58.6\arcdeg, -81.3\arcdeg)$, $(l, b) =
(267.4\arcdeg, -34.1\arcdeg)$.  Note that the two coordinate pairs
given by Baggaley do not actually correspond to the same point on the
sky.

\begin{figure}
\epsscale{1.00}
\plotone{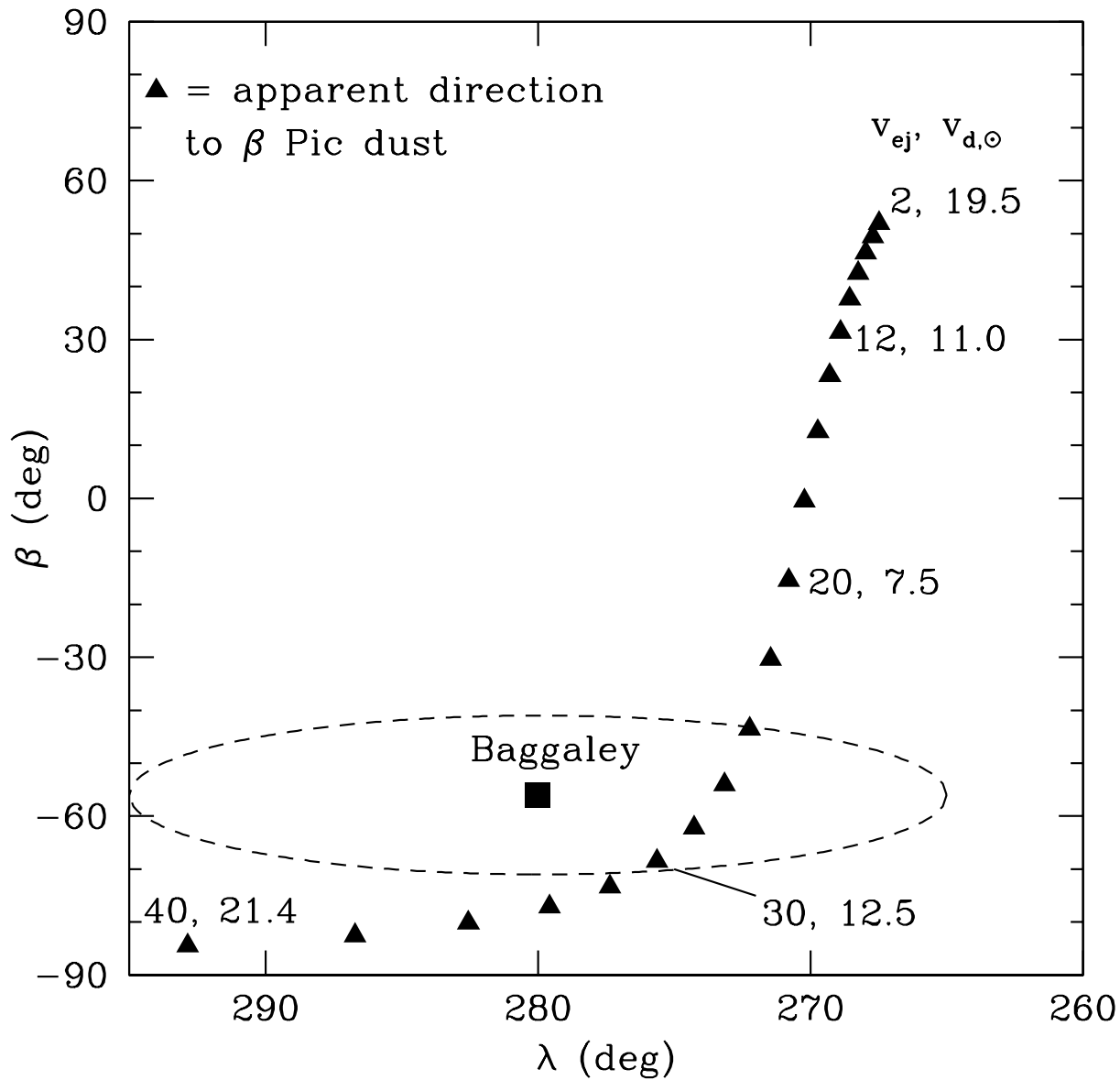}
\caption{
\label{fig:baggaley}
The box marks the central location of the discrete source observed by 
Baggaley (2000) and the dashed line indicates its extent.
Triangles mark the apparent location of a dust stream from $\beta \,$Pic for 
values of the ejection speed $\vej$ ranging from 2 to $40 \kms$ in steps of 
$2 \kms$.  For a few cases, $\vej$ and the speed of the dust with respect 
to the Sun, $v_{d, \sun}$ are indicated.  The ecliptic coordinates of 
$\beta \,$Pic itself are $(\lambda, \beta) = (82.5 \arcdeg, -74.4 \arcdeg)$.
        }
\end{figure}

In Figure \ref{fig:baggaley}, we plot
the sky position of a dust stream emitted by $\beta \,$Pic for various
values of $\vej$ (triangles).  The box indicates the central location of
Baggaley's discrete source.  For reasonable ejection speeds 
($\vej \ltsim 3 \kms$), the location of the dust stream differs dramatically
from that of the discrete source.  For $\vej \approx 30 \kms$, the locations
are much closer and $v_{d, \sun} \approx 13 \kms$.  Although the dust 
stream does lie within the $\sim 30 \arcdeg$ extent of the source in this
case, it is still $12 \arcdeg$ away from the center of the source.  

The particle flux coming from the discrete source is several orders of
magnitude larger than we would expect from a debris disk at 20
parsecs. From eqn. (\ref{eq:flux_at_earth}) we find that Beta Pic
produces a flux of $a=10\micron$ particles of $10^{-3}$ per square
kilometer per year at Earth. Given our estimate of the collecting area
of the AMOR detector, somewhat less than $10\km^2$, AMOR is not
capable of detecting such a low flux.

Thus, it appears that Baggaley's discrete source is not related to 
$\beta \,$Pic.

\subsection{Potential Distributed Sources}

In addition to the discrete source, Figures 2b and c in Baggaley
(2000) suggest the presence of a distributed, band-like feature.
Here we discuss three possible sources for a distributed feature,
namely the Galactic plane, Gould's Belt, and the spiraling of grains
in the local magnetic field.

\subsubsection{The Galactic Plane}

The scale height of young stars above the Galactic plane is 
$\sim 90 \pc$ (Gilmore \& Reid 1983).  Thus, we might expect to see a 
signature of the plane in the dust flux since we can detect grains from 
sources beyond $100 \pc$ if $a$ and $v$ are large enough (but $v$ must 
not be so large that the grains are destroyed).  The effects
of the Galactic gravitational potential cannot be ignored for such large
distances, but we do not expect the Galactic potential to deflect grains
out of the plane. 
As mentioned in \S \ref{sec:Gliese}, 
the apparent direction of a dust stream with a low ejection speed is
primarily determined by the velocity of the source with respect to the 
Sun.  Since the Galactic plane is a feature in physical space rather than
velocity space, it may seem that it should not actually appear as a band
on the sky in a dust flux map.  However, if the radial velocity of a 
source (with respect to the Sun) exceeds the dust ejection speed, then,
unless the source passed near the current location of the Sun at some time 
in the past, the dust will never reach the Sun.  Thus, though the velocity 
of the source with respect to the Sun determines the direction of the dust 
stream, the location of the source does restrict which velocities can 
give rise to an observable dust stream.  

From equations (\ref{eq:flux_at_earth}) and (\ref{eq:a_cr}),
the dust flux $F \propto d^{-3.25}$, so it is not clear that distant 
stars in the plane can indeed produce a noticeable band on top of the 
relatively isotropic distribution produced by nearby stars.  The 
following simple estimate shows that a band is expected, although at 
a low contrast level.  For stars with distance $R < 90 \pc$, the volume
element (in which young stars are found) is $dV = 4 \pi R^2 dR$, while 
$dV = (90 \pc) 2 \pi R dR$ for $R > 90 \pc$.  Thus, the ratio of the 
flux $F_{\rm plane}$ due to the Galactic plane to the isotropic flux 
$F_{\rm iso}$ (due to nearby stars) is 
\be
\frac{F_{\rm plane}}{F_{\rm iso}} \approx \frac{(180 \pc)\pi 
\int_{90 \pc}^{\infty} dR R^{-1.25}}{4 \pi \int_0^{90 \pc} dR
R^{-0.25}} = 1.6~~~,
\ee
where we have multiplied each integrand by a factor $R$ to account for 
the increased probability of catching a star when $T_{\rm ej} \approx
T_{\rm cr}$.

Since the components of random stellar velocities perpendicular to the 
plane are smaller than those parallel to the plane,
the Galactic plane should appear as a fuzzy band on
the sky even considering only stars closer than $90 \pc$.  

Note that the observed dust flux band
should be warped in such a way as to favor the direction of the solar 
apex.  

\subsubsection{Gould's Belt}

Gould's Belt is a band on the sky, inclined $\approx 20\arcdeg$ with 
respect to the Galactic plane and with an ascending node $l_{\Omega} \approx
280\arcdeg$, along which young stars tend to lie
(see P\"{o}ppel 1997 and de Zeeuw et al.~2001 for reviews).
From a Hipparcos study of early-type stars, Torra, Fern\'{a}ndez, \& 
Figueras (2000) concluded that $\approx 60\%$ of the stars younger than
$60 \Myr$ and within $600 \pc$ of the Sun are in Gould's Belt.  
OB associations, young star clusters, and molecular clouds have long been
known to trace Gould's Belt.  Recently, Guillout et al.~(1998) detected 
late-type stellar members of the Gould Belt population, by 
cross-correlating the ROSAT All-Sky Survey with the Tycho catalog.  
Thus, Gould's Belt could be responsible for a distributed grain flux 
feature.  

Guillout et al.~(1998) found that the three-dimensional structure of 
Gould's Belt is disk-like.  The outer rim of the Gould Disk is 
ellipsoidal, with a semi-major axis of $\approx 500 \pc$ and a semi-minor
axis of $\approx 340 \pc$.  The center of the structure is $\approx 200 \pc$
from the Sun, towards $l \approx 130\arcdeg$.  On the near side
($l \approx 310\arcdeg$), the inner edge of the Gould Disk lies only 
$\approx 30 \pc$ from the Sun, although most of the young stars lie 
beyond $\approx 80 \pc$.  

Estimates of the age of Gould's Belt range from 20--90$\, \Myr$
(see Torra et al.~2000 for a review); Torra et al.~favor an age between
30 and 60$\, \Myr$.  Grains traveling with $\vej = 1.5 \kms$ reach a 
distance of $\approx 90 \pc$ from their star of origin in $60 \Myr$.  
Thus, only the near side of the Gould Disk should be visible in the 
dust flux.  

The kinematics of Gould's Belt is still poorly understood, but it is clear
that the member stars are undergoing some sort of expansion.  If the 
expansion dominates the random velocities, then the near side of Gould's 
Belt should appear on the opposite side of the sky in the dust flux,
i.e., centered on $l \approx 130\arcdeg$.  If the random velocities are
more important, then we would expect to see a feature lying in the same
direction as the near side of Gould's Belt, due to stars moving towards
us.

\subsubsection{Spiraling of Grains in the Local Magnetic Field}

Thus far, we have only considered grains that are not dramatically 
deflected as they travel from their source to the Earth.  Suppose the 
magnetic field were uniform throughout the Local Bubble.  In this case,
deflected grains would produce a wide band centered on the plane 
perpendicular to the field.  The actual field presumably includes a 
significant random component, so the resulting distributed feature (if 
any) may be more complicated than a wide band.

\section{\label{sec:AGB} DUST FROM AGB STARS}

Stars with main sequence mass $M \ltsim 6 M_{\sun}$ spend time on the 
asymptotic giant branch (AGB).  During this evolutionary phase, a wind is 
driven off the star, with mass loss rates as high as $\sim 10^{-5} 
M_{\sun} \yr^{-1}$.  Grain formation in these outflows likely supplies a 
large fraction of the dust in the ISM.  See Habing (1996) for a general
review of AGB stars.

Since the grains spend only a limited time in a high-density environment,
the size distribution of the newly formed dust is thought to be dominated
by grains with $a < 1 \micron$ (see, e.g., Kr\"{u}ger \& Sedlmayr 1997).
Observations confirm the significant presence of submicron grains.  
Jura (1996) examined the extinction due to circumstellar dust around seven
O-rich giants and found that it rises towards the ultraviolet.  
From observations of scattering, Groenewegen (1997) found a mean grain size 
$a \approx 0.16 \micron$ for the circumstellar envelope of the carbon star
IRC +10 216.  

Even if submicron grains dominate the outflows from AGB stars, the 
luminosity of grains with $a \gtsim 10 \micron$ may be high enough to result
in detectable fluxes at Earth, since the dust production rate is so high.
Grains with isotopic ratios indicative of formation in AGB stars have 
been discovered in meteorites.  Although most of these grains have 
$a \sim 0.5 \micron$, some are as large as $a \approx 10 \micron$
(Zinner 1998).

Suppose the grains all have radius $a$ and that a total dust mass $M_d$ 
is emitted during the AGB lifetime $\tau$.  The dust luminosity is then
$L = 3 M_d / (4 \pi \rho a^3 \tau)$.  We adopt $M_d \sim 0.01 M_{\sun}$
and the canonical lifetime $\tau \sim 10^5 \yr$ (Habing 1996).  The flux
at Earth follows from equation (\ref{eq:specific_flux}).  For simplicity,
we take $f_{\rm beam}=1$ and $f_{\rm survive} =1$; also, since the typical
grain ejection speed in an AGB wind is $\sim 10 \kms$ (Habing 1996), 
we take $v_{d, \sun}/v_{ej} = 1$.  Thus, the flux as a function of AGB star
distance $d$ is approximately
\be
F(d) \sim 114 \yr^{-1} \km^{-2} \left( \frac{M_d}{10^{-2} M_{\sun}} \right)
\left( \frac{\rho}{3.5 \g \cm^{-3}} \right)^{-1} \left( \frac{a}{10 \micron}
\right)^{-3} \left( \frac{\tau}{10^5 \yr} \right)^{-1} \left( \frac{d}
{100 \pc} \right)^{-2}~~~.
\ee

In order for the grains to be traceable to their point of origin, we require
that the gyroradius $r_B > d$.  This yields a minimum acceptable grain size 
(eq.~\ref{eq:gyroradius}), so that 
\begin{eqnarray}
\label{eq:flux_AGB}
F(d) & \ltsim 8 \yr^{-1} \km^{-2} & \left( \frac{M_d}{10^{-2} M_{\sun}} 
\right)
\left( \frac{\rho}{3.5 \g \cm^{-3}} \right)^{1/2} \left( \frac{\Phi}{0.5 \,
{\rm V}} \right)^{-3/2} \left( \frac{B}{5 \mu {\rm G}} \right)^{-3/2} \times 
\nonumber \\
& & \left( \frac{v}{10 \kms} \right)^{3/2} 
\left( \frac{\tau}{10^5 \yr} 
\right)^{-1} \left( \frac{d} {100 \pc} \right)^{-7/2}~~~.
\end{eqnarray}
The above flux is for the optimistic case that all of the dust mass is
in the smallest grain size $a_t$ ($25\micron$ in this case) such that
the grain will barely be undeflected. In cgs units, this flux is
$2.5\times10^{-17}\cm^{-2}\s^{-1}$ tantalizingly close to the flux of
the point source seen by Baggaley (2001).

Jackson, Ivezi\'{c}, \& Knapp (2002) have studied the distribution of AGB
stars with mass loss rates in the range $10^{-6}$--$10^{-5} M_{\sun} 
\yr^{-1}$.  They find that the number density of AGB stars 
$n_{\rm AGB} \approx 4.4 \times 10^{-7} \pc^{-3}$ in the solar neighborhood.
Thus, the number of sources within a distance $d$ from the Sun is 
$N(d) \approx 1.84 (d/100 \pc)^3$.  

In Table \ref{tab:AGB} we give the flux $F$ from equation 
(\ref{eq:flux_AGB}) and the number $N$ of sources for a few values of $d$.
We also give the threshold grain size $a_t$ and the fraction $g$ of the 
emitted dust mass that would need to be in such large grains in order to 
yield an observed flux at Earth of $2 \times 10^{-3} \yr^{-1} \km^{-2}$
(corresponding to 20 events per year with a collecting area of 
$10^4 \km^2$).  We see from the table that AGB stars could be a significant
source population for radar detection of extra-solar meteoroids if a small 
fraction of the grains in the outflow are large.
Note that the source AGB star will not be visible as such, due to 
the large distances to AGB stars and their short lifetimes.  Also, the 
distances are large enough that the grain trajectories will be affected
by the Galactic potential. 

Even if all the grains in the AGB outflow are smaller than $10 \micron$,
there is still a chance to see the AGB star as a source on the sky, since
some AGB stars apparently have long-lived disks in which grain coagulation
occurs.  This phenomenon has been observed for binary stars; the presence
of the companion apparently causes some of the outflowing grains to be 
deflected into a disk.  A well-studied example is AC Her (Jura, Chen, \& 
Werner 2000).  Jura et al.~find that the grains must have $a \gtsim 200
\micron$ in order to remain gravitationally bound, due to the extreme 
luminosity of the evolved star.  They estimate the mass of the dust 
disk (in grains with $a \ltsim 0.1 \cm$) to be $\approx 1.2 \times
10^{30} \g$.  Their observations of infrared emission
imply the presence of much smaller grains as well.  They suggest that 
shattering during collisions between disk grains leads to a radiation 
pressure-driven outflow of smaller grains.  The lifetime of such disks,
and the frequency with which they occur, remain unknown.  The large disk 
mass suggests the possibility of very large dust fluxes and the commonality
of binaries suggests that the frequency could be high.  Other examples of
stars with circumbinary disks are HD 44179 (the central star of the Red 
Rectangle nebula; Jura, Turner, \& Balm 1997), IRAS 09425-6040 
(Molster et al.~2001), and SS Lep, 3 Pup, and BM Gem (Jura, Webb, \& 
Kahane 2001).  

If the AMOR point source is (or was) an AGB star, the star would
currently be a young, hot, high proper motion white dwarf. The proper
motion would point back to the vicinity of the AMOR source, after
allowing for the uncertain aberration of the dust particles.

\section{\label{sec:YSOs} DUST FROM YOUNG STELLAR OBJECTS}

The accretion of material onto a forming star is accompanied by a bipolar
outflow of material.  These outflows appear as both highly collimated,
supersonic ($v \sim 100$--$200 \kms$) jets and as large-angle molecular 
outflows (with $v \ltsim 25 \kms$).  See Reipurth \& Bally (2001), 
K\"{o}nigl \& Pudritz (2000), Shu et al.~(2000), Eisl\"{o}ffel et 
al.~(2000), and Richer et al.~(2000) for reviews.  

The mechanism for launching the outflow is not yet well understood. 
If the material originates in the disk and if the density is high enough,
then large grains can be entrained in the outflow, assuming there has been
sufficient time for growth to large sizes.  As the density decreases, 
the grains decouple from the gas and retain their launching speed, which 
could be $\sim 10$--$500 \kms$.  Chugai (2001) has appealed to the ejection
of large grains from young stellar objects (YSOs) to explain the early AMOR 
detections of interstellar meteors (Taylor, Baggaley, \& Steel
1996). He finds that the flux supplied from YSOs is a factor of 30
below that claimed by the AMOR group.

Shocks form when the jets run into the ambient ISM and grains could be  
destroyed in the shocked regions (known as Herbig-Haro objects).  
Observations have yielded contradictory results on this point.  
Beck-Winchatz, B\"{o}hm, \& Noriega-Crespo (1996) and B\"{o}hm \& 
Matt (2001) find that the gas-phase Fe abundance in Herbig-Haro objects is
generally close to solar, indicating very efficient grain destruction.
Mouri \& Taniguchi (2000), on the other hand, find the gas-phase Fe 
abundance to be $\approx 20\%$ solar, indicating only a modest amount of 
grain destruction.  Grain destruction is not expected for the slower, 
wide-angle outflows.  However, these outflows may consist of ambient 
material swept up by the jets (K\"{o}nigl \& Pudritz 2000), in which 
coagulated grains might not be present.  Thus, it is not yet clear 
whether or not YSO outflows contain large grains.

We can make an optimistic estimate of the dust luminosity by assuming
that $\sim 1\%$ of the outflowing mass is in large grains, as we did for
AGB stars.  The mass loss rate can be as high as $\sim 10^{-6} M_{\sun}
\yr^{-1}$ (Eisl\"{o}ffel et al.~2000), yielding a flux
\be 
F(d)\approx2.8\times 10^{-18}
\left(100\pc\over d\right)^2{v_{d\odot}\over v_{ej}}
\cm^{-2}\s^{-1},
\ee 
where we have assumed $a=25\micron$. For ejection in a jet, this is
$\approx 100$ times smaller than the optimistic estimate for an AGB
star at the same distance. However, the space density of YSO's is
larger than that of AGB stars; TW Hydrae, a T Tauri star, is only
$59\pc$ from the sun, and is associated with dozens of other young
stars at similar distances. There is good evidence that TW Hydrae has
a $\sim100\km\s^{-1}$ wind
\citet{2002ApJ...572..310H,2000ApJ...534L.101W}.  TW
Hydrae may be associated with Gould's Belt \citet{2001A&A...368..866M}.
While the (optimistic) flux
estimate given here is lower by a factor of order $100$ than is
realistically detectable by AMOR, TW Hydrae and stars associated with
it might be seen by more sensitive future radar systems.

The outflow may last $\sim 10^6 \yr$, 10 times longer than the AGB
star lifetime.  In the following stage of
evolution, the YSO disk is dispersed, perhaps by photoevaporation.  In
this phase, the mass loss rate could be as high as $\sim 10^{-7}
M_{\sun} \yr^{-1}$ (Hollenbach, Yorke, \& Johnstone 2000), yielding a
maximum dust luminosity 10 times lower than for the bipolar outflows.

Given the potentially large dust luminosities, YSOs could be likely 
candidate sources for extrasolar meteoroids.  However, the young stars 
listed in Table \ref{tab:Gliese} are probably too old, given their 
proximity.  For example, if disk dispersal is complete at an age of 
$3 \Myr$, then dust from the YSO phase of $\beta \,$Pic has been traveling
for $\approx 9 \Myr$.  At a speed of $10 \kms$, the dust would now be 
$\approx 90 \pc$ from $\beta \,$Pic, but the Sun is only $\approx 20 \pc$
from $\beta \,$Pic.  If, however, disk dispersal continues until an age
of $10 \Myr$, then $\beta \,$Pic may appear as a YSO dust source today.

Some nearby, young clusters could produce large total fluxes distributed
over a wide area on the sky.  Such clusters include TW Hydrae ($d \sim 55
\pc$, $t \sim 10 \Myr$; Zuckerman \& Webb 2000), Upper Scorpius 
($d \sim 145 \pc$, $t \sim \,$1--10$\Myr$; Spangler et al.~2001),
Chamaeleon Ia and Ib ($d \sim \,$140--150$\pc$, $t \sim \,$1--40$\Myr$;
Spangler et al.~2001), and Taurus ($d \sim 140 \pc$, $t \sim \,$10--40$\Myr$;
Spangler et al.~2001).

\section{DISCUSSION\label{sec:Discussion}}
Our calculations suggest that there are a number of nearby sources of
$10\micron$ or larger particles that yield sufficient fluxes to be
detectable by ground based radar systems. Here we suggest ways to
optimize radar systems for meteor detection. 

Meteoroids with $a\lesssim 10\micron$ cannot travel through the ISM for
appreciable distances. Furthermore, it appears that the flux of meteoroids
decreases with increasing meteoroid size. It follows that if one is interested
in detecting particles from an identifiable source, large radar
collecting area is more crucial than large radar power, once the power
is sufficient to detect $a\approx10\micron$ particles.

The returned radar power falls off as $1/R^3$ for radars that detect
coherent emission from a substantial fraction of the meteor trail
(such as AMOR), or as $1/R^4$ for meteor head detectors such as
Arecibo. The minimum range to the meteor trail is given by the height
of the trails (of order $100\km$), while the maximum range is of order
$1000\km$ due to the curvature of the Earth. Increasing the mean
range by a factor of 10 will increase the collecting area, and hence
the flux, by a factor of 100, but requires an increase in radar power
by a factor of $10^3$ to $10^4$, depending on the type of radar
employed. While the cost of a high power radar transmitter increases
rapidly with increasing power, it is clearly helpful to maximize the
transmitted power.

Another way to increase the radar range, and hence the collecting
area, is to increase the gain of the antenna; the received power
scales as the product $G_TG_R$. However, there are limits to the
extent to which one can increase the antenna gain. If the radar is
like AMOR, the width of the beam must exceed the length of a typical
meteor trail, else the number of coherently emitting electrons will
drop. The angular width of the trail is of order $H_p/R$, which ranges
from $0.06$ to $0.006$ radians, or $3^\circ$ (in RA, if the beam looks
due south or north) at the zenith to $0.3^\circ$ at the
horizon. Optimally, the beam will cover the sky from the zenith to the
horizon, so the antenna gain should be no larger than $\sim 1000$.

Selecting a gain of this magnitude matches the beam size to the length
of the meteor trail, but limits the geometric area $A_G$ of the
antenna, partially defeating the purpose, which is to maximize the
collecting area. This is the case with AMOR, where the choice of a
narrow beam was dictated by the need for high precision measurement of
the meteor position, but at the cost of collecting area. To get around
this problem, we propose building an array of antennas, all powered by
the same radio generator. The beams of the array would all have widths
of order $3^\circ$, but the beams would be directed around the points
of the compass. In principle one could have $\sim100$ such beams
emanating from a single facility. The duration of a radar pulse is
typically a microsecond, while the time between pulses is of order a
millisecond. (The time to travel out and back $1000\km$ is about 6
milliseconds; in that time the meteoroid will travel a distance
$\sim240\m$.) One could send out pulses, directed at different points
around the compass, separated by 10 milliseconds. The receiver would
be turned off when the transmitter is on, about $10^{-4}$ of the
time. Return pulses from trails at different compass points would
overlap in time, but given sufficient receiver gains, cross talk
between the receivers should be small enough to be acceptable.

If one employs a sufficiently powerful radar, one can live with a low
antenna gain and still achieve large collecting area. However, the need for
high precision determinations of the meteor positions remains. This
problem can be addressed by using separate transmitting and receiving
antennae. The transmitter can have a broad beam, with a low gain but
covering a large fraction of the sky. An antenna array can then be used as
an interferometer to locate the meteor on the sky.

Either a multi-beam radar or single broad beam radar with
an interferometer could have a collecting area
$A_{col}$ of order $800\km^2$, if the typical range were $R\sim
150\km$. Current distant early warning (DEW) radar may be interesting
in this context. The existence of tens of such systems, with an
aggregate collecting area of $\sim10,000\km^2$ only adds to their
attractiveness as extrasolar meteor detectors.

The collecting area grows dramatically, up to $35,000\km^2$,
if the radar can detect $10\micron$ particles out to the
horizon. This is $4000$ times the collecting area of the AMOR
radar. AMOR sees about 10 extrasolar meteors every day; the proposed
detector would see up to $40,000$ extrasolar meteoroids of size
$35\micron$ a day, and about $10^6$ $10\micron$ extrasolar
meteoroids per day. This suggests that one of the major constraints is
handling the data, since there would be of order $10^9$ interplanetary
dust particles detected per day.

Such a radar would be capable of detecting meteoroids from a handful of
nearby debris disks; table \ref{tab:Gliese} lists four nearby young
Gliese stars that might have fluxes exceeding $6\times10^{-4}$
particles per year per square kilometer. This large area radar system
would also detect meteoroids from (former) AGB stars with fluxes as low as
$2\times10^{-11}\cm^{-2}\s^{-1}$; in the optimistic case that a
substantial fraction of emitted particles have $a\gtsim25\micron$,
sources as distant as a kiloparsec could be seen. From table
\ref{tab:AGB} this could be several hundred stars. The number of
possible YSO sources could be somewhere between these two estimates.

How feasible is such a radar? We have already mentioned the current DEW
systems. However, we note that moderate enhancements to an
AMOR type radar are all that are needed. The product of the
transmitter and receiver gain could be enhanced by a factor of $5-10$,
while the power could be enhanced by a factor of 20, to $2$ megawatts,
for a total increase in received power of a factor of $\sim200$. This
would allow for the detection of $a=10\micron$ size particles at
ranges of $\sim500\km$. Further progress could be made by increasing
the operating wavelength; a factor of two increase, to
$\lambda=2000\cm$ would be sufficient to detect $10\micron$ size
particles out to the horizon.

The flux of interstellar meteoroids of different mass appears to follow a
power law $mf_m\propto m^{-1.1}$, over five decades in mass. This may
be compared to the Dohnanyi law, $f_m\propto m^{-\alpha}$, with
$\alpha=11/6\approx 1.83$; in the case of interstellar meteoroids,
$\alpha\approx2.1$. We note that the observed size distribution of
asteroids has $\alpha\approx2$ for objects larger than $a=2.5\km$, and
$\alpha\approx 1.4$ for $a$ between $2.5$ and $0.1\km$
\cite{Ivezic}. However, the binding energy per gram can be expected to
vary with size, a fact we appealed to in explaining the observation
that meteoroids of very difference sizes ablate at the same height in the
Earth's atmosphere. In our view, the rough agreement between the
values of $\alpha$ might reflect a common origin in a collisional
cascade, but is more likely a coincidence.

Krivova \& Solanki \citep{2003A&A...402L...5K} recently suggested that
interactions with a Jupiter mass planet could eject dust grains with
velocities of order $10-70\km\s^{-1}$. They cited this result as
support for the identification of the point source found by Baggaley
(2000) with $\beta\,$ Pic. Our analytic calculations show that only a
tiny fraction of ejected particles will have such high velocities, a
result supported by direct numerical integrations.

\section{CONCLUSIONS\label{sec:conclusions}}

We have provided simple analytic estimates for the minimum size of
radar detected meteoroids as a function of radar power and antenna
gain. We give very rough estimates of the collecting area of radar
detectors, and hence of fluxes of interstellar meteoroids, assuming that
the claimed detection rates are correct. The fluxes of satellite and radar
detected interstellar meteoroids appear to lie along a power law,
$mf_m\propto m^{-1.1}$, corresponding to a $\alpha=2.1$, or a size
scaling $\gamma=4.3$, in contrast to the Dohnanyi value $\gamma=3.5$.

We examine three possible sources of large, $a>10\micron$ size
extrasolar meteoroids. In descending order of dust luminosity, they are
AGB stars, young stellar objects (with winds and/or jets), and debris
disks. (Supernovae can supply similar mass fluxes of particles, but we
expect the particles to be smaller.) We show that such large particles
can travel in straight lines for tens of parsecs through the
interstellar medium, which in principle will allow their sources to be
traced. 

Since there are $\sim2$ AGB stars within 100pc of the sun at any time,
we expect that there will be $\sim2$ sources of extrasolar meteoroids,
with fluxes at Earth approaching $\sim8\yr^{-1}\km^{-2}$, on the sky
at all times. Such a star might be responsible for the ``point
source'' reported in Baggaley (2000). The AMOR radar, according to our
estimates, is marginally capable of detecting such a small flux. The
star itself would currently be a hot white dwarf with a high proper
motion, within $\sim100\pc$ of the sun.

There is one known T Tauri star within $100\pc$ of the sun, TW
Hydrae. This star has a high velocity outflow ($\sim100\km\s^{-1}$)
from a dusty disk. It is very likely expelling $10$ to $100\micron$
size particles; unfortunately, the flux is clearly too small, by a
factor of $\sim100$, to be detected by current radar systems.

Similarly, currently known debris disk systems provide fluxes that are
too small to be detected at present. We have shown that the system
implicated by Baggaley (2000), $\beta$ Pic, is unlikely to provide a
particle flux as large as that associated with his point source. We
also show that the location of the apparent source on the sky is
inconsistent with particles ejected from $\beta$ Pic by gravitational
interaction with a Jupiter mass planet in the system; we showed that
typical ejection velocities in that case are of order
$1\km\s^{-1}$. Even if the ejection velocity is assumed to be
$30\km\s^{-1}$, the most favorable value, the apparent position of the
source does not match the observed position (see
Fig. \ref{fig:baggaley}). Since the particle flux, ejection velocity,
and position on the sky do not match the observations, we conclude
that the point source is not associated with gravitational ejection
from the debris disk of $\beta$ Pic.

However, we would like to stress that all three types of sources, AGB
stars, YSOs, and debris disks, should be detectable by future ground based
radar systems. The pioneering AMOR system may have already detected
meteoroids from an AGB star. Modest improvements in this type of ground
based radar system, in particular better sky coverage, should allow
for the detection of multiple examples of each type of source. An
alternate possibility is to piggyback on already existing radar
employed in distant early warning systems.

\acknowledgements 
We are grateful to C.~D.~Matzner, D.~N.~Spergel, D.~D. Meisel, W.~J.~Baggaley,
B.~T.~Draine, and R.~E.~Pudritz for helpful discussions.  This research
was supported by NSERC of Canada and has made use of the SIMBAD
database, operated at CDS, Strasbourg, France, and of NASA's
Astrophysics Data System.

\appendix
\section{\bf APPENDIX}
In this appendix we rederive some of the relations used in section
\ref{sec:detectable_fluxes}, employing the classical theory of
meteors.  In the meteoroid frame, the air flow has a kinetic energy flux
\be 
F={1\over2}\rho_av^3.
\ee 
Following the meteoritics literature, we define a shape factor $A$ to
be the ratio of the area $A_m$ of the meteoroid divided by the two-thirds
power of the volume $(m/\rho)$, or 
\be 
A\equiv{A_m\over(m/\rho)^{2/3}}.
\ee 
For a sphere, $A=(9\pi/16)^{1/3}\approx 1.2$. The kinetic luminosity
seen by the meteoroid is then $A_m F= {A\over2} (m/\rho)^{2/3}\rho_av^3$; assume
that a fraction $\Lambda$ of this kinetic
luminosity goes toward ablating the meteor. Then the time rate
of change of the binding energy is
\be 
{dE\over dt}=-{\Lambda A\over2}\left({m\over\rho}\right)^{2/3}\cdot\rho_av^3.
\ee 
Typical estimates for $\Lambda$ are around $0.5$. 
It follows that the rate of ablation is 
\be \label{ablation}
{dm\over dt}=-{\Lambda A\over 2\zeta}
\left({m\over \rho}\right)^{2/3}\rho_av^3,
\ee 
where, the reader will recall, the binding energy $E$ is related to
the meteoritic mass $m$ by the heat of ablation $\zeta$, $E=\zeta m$.
From equation (\ref{q}), the number of ions produced per
centimeter along the path is
\be 
q=-{\beta\over v\mu}{dm\over dt},
\ee 
where $\mu$ is the mean molecular weight of the meteoroid (recall that
$\beta$ is the number of ions produced by each meteoroid atom). Combining
this with equation (\ref{ablation}), we find 
\be  
q=\beta{\Lambda A\over 2\zeta\mu}
\left({m\over \rho}\right)^{2/3}\rho_av^2.
\ee  
This is known as the ionization equation.

The maximum line density along the track can be found by
differentiating the ionization equation with respect to time:
\be  \label{dqdt}
{1\over q}{dq\over dt}={1\over\beta}{d\beta(v)\over
dt}+{2\over3m}{dm\over dt}+{1\over\rho_a}{d\rho_a\over dt}+ {2\over
v}{dv\over dt}.
\ee  
Note that $\beta(v)\sim v^n$, so that both the first and last terms on
the right hand side are proportional to 
\be  
{1\over v}{dv\over dt}
\ee  
We assume for the moment that this is  smaller than either the
mass or density variations. Then the maximum value of $q$ occurs when
\be  
{2\over 3m}{dm\over dt}=-{1\over \rho_a}{d\rho_a\over dt},
\ee  
or
\be  
{2\over3m}(-{q_{max}\mu\over \beta}v)=-{v\over H_p},
\ee  
where $H_p$ is the density scale height. Thus
\be  
q_{max}={3\over2}{\beta m\over\mu H_p}.
\ee  
In words, the maximum line density is given by spreading the meteoroid over
approximately a scale height and accounting for the number $\beta$ of
atoms that are ionized for each atom in the meteor.

Next we justify the neglect of terms proportional to the derivative of
the velocity.  The momentum equation for the meteoroid is
\be 
m{dv\over dt}=mg-C_DA\left({m\over\rho}\right)^{2/3}\rho_av^2,
\ee 
where $C_D$ is the drag coefficient. The mean free path at meteor
heights is of order $10\cm$, much larger than the typical radius of
the meteoroids we are interested in. In that case $C_D=1$. The velocity
derivative becomes
\be  \label{dvdt}
{1\over v}{dv\over dt}={g\over v}-A\left({m\over\rho}\right)^{2/3}\rho_av/m
\ee  
We note that $H_p\approx c^2/g$ where $c$ is the sound speed; then 
\be  
{v\over H_p}\approx\left({v\over c}\right)^2{g\over v}>>{g\over v}
\ee  
so that the first term on the right hand side of (\ref{dvdt}) is
negligible compared to $v/H_p$; recall that the meteoroids are highly
supersonic. Next we compare the velocity gradient in (\ref{dqdt})
(including only the second term on the rhs of
(\ref{dvdt})),
\be  
{2+n\over v}{dv\over dt}\approx A\left({m\over\rho}\right)^{2/3}
{\rho_av\over m}\sim a^2\rho_av/\rho
a^3\sim{\rho_a\over \rho}{H_p\over a}{v\over H_p}
\ee  
to
\be  
{2\over 3m}{dm\over dt}=-{\Lambda A\over2\zeta}
\left({m\over\rho}\right)^{2/3}
{\rho_a v^3\over m}.
\ee  
The ratio is
\be  
{12\over \Lambda}{\zeta\over v^2}
={3\over4}\left({\zeta\over10^{11}\cm^2\s^{-2}}\right)
\left({v\over 40\km\s^{-1}}\right)^2
\left({10^{-1}\over\Lambda}\right),
\ee  
where we took $n=2$.

McKinley gives estimates for $\Lambda$ that are of order unity. For
these values, the velocity derivative is much smaller than the mass
derivative. We conclude that the logarithmic mass derivative is
comparable to or larger than the logarithmic velocity derivative.

\label{Table:radar}
\begin{deluxetable}{rrr} 
\tablecolumns{3} 
\tablewidth{0pc} 
\tablecaption{Radar Parameters} 
\tablehead{ 
\colhead{}    &  \colhead{Arecibo} &   \colhead{AMOR}  
}
\startdata 
$P_T$	&	2MW			&	100kW			\\
$P_n$	&	$10^{-8}{\rm \,erg/s}$&	$1.6\times10^{-7}{\rm \,erg/s}$\\
$G_T$	&	$10^6$			&	$430$			\\
$G_R$	&	$10^6$			&	$130$			\\
$R_0$	&	$100{\,\rm km}$		&	$230{\,\rm km}$		\\
$\lambda$&	$70{\,\rm cm}$		&	$1145{\,\rm cm}$\\
Beam dimensions&   ${1\over 6}^\circ\times{1\over 6}^\circ$      &
$3^\circ\times18^\circ$\\
 $A_G$ & $0.9\km^2$      & $50\km^2$              \\
 $A_{col}$ & $10^{-2}\km^2$      & $8\km^2$               \\
\cutinhead{meteor properties}
$q_{min}$&	$10^7{\,\rm cm}^{-1}$	&$7\times10^8{\,\rm cm}^{-1}$	\\
$m_{min}$&	$3\times10^{-9}{\,\rm gm}$&	$2\times10^{-7}{\,\rm gm}$	\\
$a_{min}$&	$6{\,\rm \micron}$	&	$25{\,\rm \micron}$	\\

\enddata 
\end{deluxetable} 



\begin{deluxetable}{cccc}
\tablewidth{0pc}
\tablecaption{Idealized Interstellar Environments
\label{tab:phases}}
\tablehead{
\colhead{Phase}&
\colhead{$T_{\rm gas}$}&
\colhead{$\nH$}&
\colhead{$x_e$}
\\
\colhead{}&
\colhead{$\K$}&
\colhead{$\cm^{-3}$}&
\colhead{}
}
\startdata
CNM	&100	&30	&$1.5 \times 10^{-3}$\\
WNM	&6000	&0.3	&0.1\\
LB	&$10^6$ &$5\times 10^{-3}$ 	&1\\
\enddata
\end{deluxetable}

\begin{deluxetable}{cccc}
\tablewidth{0pc}
\tablecaption{Threshold Distances for Observing Vega-Like Stars
\label{tab:d_th}}
\tablehead{
\colhead{$A_{col}$}&
\colhead{$t_{\rm cr}$}&
\colhead{$d_{\rm th}$}&
\colhead{$a_{\rm cr}$}
\\
\colhead{$\km^2$}&
\colhead{$\Myr$}&
\colhead{$\pc$}&
\colhead{$\micron$}
}
\startdata
$10^4$	&3	&26	&17\\
$10^4$	&10	&8.6	&10\\
$10^6$	&3	&108	&35\\
$10^6$	&10	&36	&20\\
\enddata
\end{deluxetable}

\begin{deluxetable}{cccccccccccc}
\tablewidth{0pc}
\tablecaption{Gliese Star Dust Fluxes
\label{tab:Gliese}}
\tablehead{
\colhead{Gl \#}&
\colhead{Name}&
\colhead{$d$}&
\colhead{$t$\tablenotemark{a}}&
\colhead{$U$\tablenotemark{b}}&
\colhead{$V$\tablenotemark{b}}&
\colhead{$W$\tablenotemark{b}}&
\colhead{$v_{\ast, \sun}$}&
\colhead{$a_{\rm cr}$}&
\colhead{$F$\tablenotemark{c}}&
\colhead{$\lambda$\tablenotemark{d}}&
\colhead{$\beta$\tablenotemark{d}}
\\
\colhead{}&
\colhead{}&
\colhead{pc}&
\colhead{Myr}&
\colhead{$\kms$}&
\colhead{$\kms$}&
\colhead{$\kms$}&
\colhead{$\kms$}&
\colhead{$\micron$}&
\colhead{$\yr^{-1} \km^{-2}$}&
\colhead{deg}&
\colhead{deg}
}
\startdata
68.0  &DM+19 279         &7.5   &69.8   &34.5   &-24.7  &2.4   
&42.5   &4.4	&3.3E-5  &62   &33\\

71.0  &$\tau \,$Cet      &3.6	&170 	&18.6	&29.4	&12.7	
&37.0	&3.3	&6.3E-6  &94   &-54\\

111.0 &$\tau^1 \,$Eri    &14.0  &39.6   &-26.2  &-16.9  &-13.0  
&33.8   &6.8	&1.5E-4  &262  &33\\

121.0 &$\tau^3 \,$Eri    &26.4  &215    &19.4   &9.96   &-1.09
&21.9   &11.6	&3.4E-8  &101  &-18\\

144.0 &$\epsilon \,$Eri  &3.22  &59.7   &-3.01  &7.22   &-20.0
&21.5   &4.1	&1.1E-4  &184  &8\\

167.1 &$\gamma \,$Dor    &20.3  &107    &-22.9  &-18.1  &-13.8
&32.3   &8.4	&1.2E-6  &261  &38\\

217.1 &$\zeta \,$Lep     &21.5  &69.8   &-14.4  &-11.2  &-8.45
&20.1   &10.9	&3.7E-6  &260  &38\\

219.0 &$\beta \,$Pic     &19.3  &12.0   &-10.8  &-16.0  &-9.11
&21.4   &10.0	&1.5E-3  &268  &51\\

248.0 &$\alpha \,$Pic    &30.3  &194    &-35.4  &-19.3  &-10.1
&41.6   &9.0	&1.0E-7  &268  &25\\

297.1 &B Car             &21.4  &31.4   &19.8   &-15.1  &-34.0
&42.2   &7.5	&1.1E-3  &140  &48\\

321.3 &$\delta \,$Vel    &24.4  &351    &11.6   &-1.15  &-5.00
&12.7   &14.6	&2.3E-9  &107  &17\\

364.0 &DM-23 8646        &14.9  &69.3   &-39.9  &-25.4  &6.27
&47.7   &5.9	&1.4E-5  &290  &17\\

448.0 &$\beta \,$Leo     &11.1  &135    &-20.1  &-16.1  &-7.80
&26.9   &6.8	&1.2E-6  &268  &36\\

557.0 &$\sigma \,$ Boo   &15.5  &123    &2.06   &15.9   &-5.17
&16.9   &10.1	&5.7E-7  &165  &-38\\

673.1 &DM-24 13337       &25.7  &34.5   &-35.8  &-13.1  &-11.6
&39.9   &8.5	&6.4E-4  &261  &20\\

691.0 &$\mu \,$Ara       &15.3  &58.8   &-13.7  &-8.39  &-3.97
&16.5   &10.1	&9.7E-6  &270  &26\\

721.0 &$\alpha \,$Lyr    &7.76  &152    &-16.1  &-6.33  &-7.76
&18.9   &6.7	&1.1E-6  &256  &26\\

820.0 &61 Cyg            &3.48  &48.0   &-93.8  &-53.4  &-8.33
&108    &1.9	&1.0E-3  &278  &23\\

822.0 &$\delta \,$Eql    &18.5  &80.0   &5.72   &-28.6  &-10.3
&30.9   &8.2	&4.3E-6  &18   &78\\

881.0 &$\alpha \,$PsA    &7.69  &83.1   &-5.70  &-8.22  &-11.0
&14.9   &7.6	&5.7E-6  &235  &54\\

\enddata
\tablenotetext{a}{Age of star from eq.~(\ref{eq:T_tau})}
\tablenotetext{b}{Space velocity of the star relative to the Sun; $U$ is 
positive towards the Galactic center, $V$ is positive along the direction 
of the Galactic rotation, and $W$ is positive towards the North Galactic Pole.}
\tablenotetext{c}{Dust flux calculated using equation
(\ref{eq:flux_at_earth}) with a limiting grain size $a=10 \micron$, a
flat distribution of ejection velocities between $0.5 \kms$ and $3 \kms$,
$T_{\rm cr} = 3 \Myr$, and $f_{\rm survive} = f_{\rm beam} = f_{\rm ej}
=1$.}
\tablenotetext{d}{Apparent location of dust stream on the sky; $\lambda$ is
ecliptic longitude and $\beta$ is ecliptic latitude.  In most cases, 
$v_{\rm ej}=1 \kms$ is adopted; for Gl 219.0 ($\beta \,$Pic) 
$v_{\rm ej}=3 \kms$ is adopted, since the dust will not yet have reached us if 
$v_{\rm ej}=1 \kms$.}
\end{deluxetable}

\begin{deluxetable}{ccccc}
\tablewidth{0pc}
\tablecaption{AGB Star Dust Fluxes
\label{tab:AGB}}
\tablehead{
\colhead{$d$\tablenotemark{a}}&
\colhead{$F$\tablenotemark{b}}&
\colhead{$N$\tablenotemark{c}}&
\colhead{$a_t$\tablenotemark{d}}&
\colhead{$g$\tablenotemark{e}}
\\
\colhead{\pc}&
\colhead{$\yr^{-1} \km^{-2}$}&
\colhead{}&
\colhead{$\micron$}&
\colhead{}
}
\startdata
100	&7.99	 &1.84	&24.3	&2.50E-4\\
200	&0.706	 &14.7	&34.3	&2.83E-3\\
500	&2.86E-2 &230 	&54.2	&6.99E-2\\
\enddata
\tablenotetext{a}{distance to AGB star}
\tablenotetext{b}{flux from eq.~(\ref{eq:flux_AGB})}
\tablenotetext{c}{number of AGB stars within distance $d$ from Sun}
\tablenotetext{d}{threshold grain size for traceability to source}
\tablenotetext{e}{fraction of emitted grains that must have $a>a_t$ in 
order for flux at Earth to equal $2 \times 10^{-3} \yr^{-1} \km^{-2}$}
\end{deluxetable}

\end{document}